\documentclass[iop]{emulateapj}
\usepackage{graphicx}
\usepackage{enumerate}
\usepackage{amssymb, amsmath}
\usepackage{natbib}
\usepackage{color}
\usepackage{ulem}
\usepackage{dblfnote}
\usepackage{appendix}

\def\pasa{PASA}
\def\procspie{Proc.~SPIE}

\newcommand{\tokyo}{1}
\newcommand{\abc}{2}
\newcommand{\kogakuin}{3}
\newcommand{\chile}{4}
\newcommand{\arizona}{5}
\newcommand{\subaru}{6}
\newcommand{\ipag}{7}
\newcommand{\lablaglange}{8}
\newcommand{\Maxplanck}{9}
\newcommand{\ucsb}{10}
\newcommand{\ias}{11}
\newcommand{\charleston}{12}
\newcommand{\ucberkeley}{13}
\newcommand{\munchen}{14}
\newcommand{\gsfc}{15}
\newcommand{\eureka}{16}
\newcommand{\Gca}{17}
\newcommand{\naoj}{18}
\newcommand{\hawaii}{19}
\newcommand{\isas}{20}
\newcommand{\princeton}{21}
\newcommand{\osaka}{22}
\newcommand{\hayama}{23}
\newcommand{\hiroshima}{24}
\newcommand{\cab}{25}
\newcommand{\jpl}{26}
\newcommand{\cincinnati}{27}
\newcommand{\ssi}{28}
\newcommand{\sokendai}{29}
\newcommand{\taiwan}{30}
\newcommand{\zurich}{31}
\newcommand{\kavli}{32}
\newcommand{\hokkaido}{33}
\newcommand{\oklahoma}{34}
\newcommand{\tohoku}{35}

\begin{document}

\author{Taichi Uyama$^\tokyo$}
\author{Jun Hashimoto$^\abc$}
\author{Takayuki Muto$^\kogakuin$}
\author{Eiji Akiyama$^\chile$}
\author{Ruobing Dong$^\arizona$}
\author{Jerome de Leon$^\tokyo$}
\author{Itsuki Sakon$^\tokyo$}
\author{Tomoyuki Kudo$^{\subaru}$}
\author{Nobuhiko Kusakabe$^\abc$}
\author{Masayuki Kuzuhara$^{\abc}$}
\author{Mickael Bonnefoy$^{\ipag}$}
\author{Lyu Abe$^{\lablaglange}$}
\author{Wolfgang Brandner$^{\Maxplanck}$}
\author{Timothy D. Brandt$^{\ucsb,\ias}$}
\author{Joseph C. Carson$^{\charleston,\Maxplanck}$}
\author{Thayne Currie$^{\subaru}$}
\author{Sebastian Egner$^{\subaru}$}
\author{Markus Feldt$^{\Maxplanck}$}
\author{Jeffrey Fung$^{\ucberkeley}$}
\author{Miwa Goto$^{\munchen}$}
\author{Carol A. Grady$^{\gsfc,\eureka,\Gca}$}
\author{Olivier Guyon$^{\abc,\subaru,\arizona}$}
\author{Yutaka Hayano$^{\subaru}$}
\author{Masahiko Hayashi$^{\naoj}$}
\author{Saeko S. Hayashi$^{\subaru}$}
\author{Thomas Henning$^{\Maxplanck}$}
\author{Klaus W. Hodapp$^{\hawaii}$}
\author{Miki Ishii$^{\naoj}$}
\author{Masanori Iye$^{\naoj}$}
\author{Markus Janson$^{\princeton}$}
\author{Ryo Kandori$^{\naoj}$}
\author{Gillian R. Knapp$^{\princeton}$}
\author{Jungmi Kwon$^{\isas}$}
\author{Taro Matsuo$^{\osaka}$}
\author{Satoshi Mayama$^{\hayama}$}
\author{Michael W. Mcelwain$^{\gsfc}$}
\author{Shoken Miyama$^{\hiroshima}$}
\author{Jun-Ichi Morino$^{\naoj}$}
\author{Amaya Moro-Martin$^{\princeton,\cab}$}
\author{Tetsuo Nishimura$^{\subaru}$}
\author{Tae-Soo Pyo$^{\subaru}$}
\author{Eugene Serabyn$^{\jpl}$}
\author{Michael L. Sitko$^{\cincinnati,\ssi}$}
\author{Takuya Suenaga$^{\naoj,\sokendai}$}
\author{Hiroshi Suto$^{\abc,\naoj}$}
\author{Ryuji Suzuki$^{\naoj}$}
\author{Yasuhiro H. Takahashi$^{\tokyo,\naoj}$} 
\author{Michihiro Takami$^{\taiwan}$}
\author{Naruhisa Takato$^{\subaru}$}
\author{Hiroshi Terada$^{\naoj}$}
\author{Christian Thalmann$^{\zurich}$}
\author{Edwin L. Turner$^{\princeton,\kavli}$}
\author{Makoto Watanabe$^{\hokkaido}$}
\author{John Wisniewski$^{\oklahoma}$}
\author{Toru Yamada$^{\tohoku}$}
\author{Yi Yang$^{\sokendai}$}
\author{Hideki Takami$^{\naoj}$}
\author{Tomonori Usuda$^{\naoj}$}
\author{Motohide Tamura$^{\tokyo,\abc,\naoj}$}

\footnotetext[\tokyo]{Department of Astronomy, The University of Tokyo, 7-3-1, Hongo, Bunkyo-ku, Tokyo 113-0033, Japan}
\footnotetext[\abc]{Astrobiology Center of NINS, 2-21-1, Osawa, Mitaka, Tokyo 181-8588, Japan}
\footnotetext[\kogakuin]{Division of Liberal Arts, Kogakuin University, 1-24-2, Nishi-Shinjuku, Shinjuku-ku, Tokyo, 163-8677, Japan}
\footnotetext[\chile]{Chile Observatory, National Astronomical Observatory of Japan, 2-21-2, Osawa, Mitaka, Tokyo, 181-8588, Japan}
\footnotetext[\arizona]{Steward Observatory, University of Arizona, Tucson, AZ 85721, USA}
\footnotetext[\subaru]{National Astronomical Observatory of Japan, Subaru Telescope, National 
Institutes of Natural Sciences, Hilo, HI 96720, USA}
\footnotetext[\ipag]{Univ. Grenoble Alpes, CNRS, IPAG, F-38000 Grenoble, France}
\footnotetext[\lablaglange]{Laboratoire Lagrange (UMR 7293), Universite de Nice-Sophia Antipolis, CNRS, Observatoire de la Coted'azur, 28 avenue Valrose, 06108 Nice Cedex 2, France}
\footnotetext[\Maxplanck]{Max Planck Institute for Astronomy, K$\ddot{o}$nigstuhl 17, 69117 Heidelberg, Germany}
\footnotetext[\ucsb]{Department of Physics, University of California-Santa Bar-bara, Santa Barbara, CA, USA}
\footnotetext[\ias]{Astrophysics Department, Institute for Advanced Study, Princeton, NJ, USA}
\footnotetext[\charleston]{Department of Physics and Astronomy, College of Charleston, 66 George St., Charleston, SC 29424, USA}
\footnotetext[\ucberkeley]{Department of Astronomy, University of California, Berkeley, CA, USA}
\footnotetext[\munchen]{Universit$\ddot{a}$ts-Sternwarte M$\ddot{u}$nchen, Ludwig-Maximilians-Universit$\ddot{a}$t, Scheinerstr. 1, 81679 M$\ddot{u}$nchen,Germany}
\footnotetext[\gsfc]{Exoplanets and Stellar Astrophysics Laboratory, Code 667, Goddard Space Flight Center, Greenbelt, MD 20771, USA}
\footnotetext[\eureka]{Eureka Scientific, 2452 Delmer, Suite 100, Oakland CA96002, USA}
\footnotetext[\Gca]{Goddard Center for Astrobiology}
\footnotetext[\naoj]{National Astronomical Observatory of Japan, 2-21-1, Osawa, Mitaka, Tokyo, 181-8588, Japan}
\footnotetext[\hawaii]{Institute for Astronomy, University of Hawaii, 640 N. A‘ohoku Place, Hilo, HI 96720, USA}
\footnotetext[\isas]{Institute of Space and Astronautical Science, JAXA, 3-1-1 Yoshinodai, Sagamihara, Kanagawa Japan}
\footnotetext[\princeton]{Department of Astrophysical Science, Princeton University, Peyton Hall, Ivy Lane, Princeton, NJ08544, USA}
\footnotetext[\osaka]{Department of Earth and Space Science, Graduate School of Science, Osaka University, 1-1 Machikaneyamacho, Toyonaka, Osaka 560-0043, Japan}
\footnotetext[\hayama]{SOKENDAI(The Graduate University for Advanced Studies), Shonan International Village, Hayama-cho, Miura-gun, Kanagawa 240-0193, Japan
}
\footnotetext[\hiroshima]{Hiroshima University, 1-3-2, Kagamiyama, Higashihiroshima, Hiroshima 739-8511, Japan}
\footnotetext[\cab]{Department of Astrophysics, CAB-CSIC/INTA, 28850 Torrej$\acute{o}$n de Ardoz, Madrid, Spain}
\footnotetext[\jpl]{Jet Propulsion Laboratory, California Institute of Technology, Pasadena, CA, 171-113, USA}
\footnotetext[\cincinnati]{Department of Physics, University of Cincinnati, Cincinnati, OH 45221, USA}
\footnotetext[\ssi]{Center for Extrasolar Planetary Systems, Space Science Institute, 4750 Walnut St, Suite 205, Boulder, CO 80301, USA}
\footnotetext[\sokendai]{Department of Astronomical Science, The Graduate University for Advanced Studies, 2-21-1, Osawa, Mitaka, Tokyo, 181-8588, Japan}
\footnotetext[\taiwan]{Institute of Astronomy and Astrophysics, Academia Sinica, P.O. Box 23-141, Taipei 10617, Taiwan}
\footnotetext[\zurich]{Swiss Federal Institute of Technology (ETH Zurich), Institute for Astronomy, Wolfgang-Pauli-Strasse 27, CH-8093 Zurich, Switzerland}
\footnotetext[\kavli]{Kavli Institute for Physics and Mathematics of the Universe, The University of Tokyo, 5-1-5, Kashiwanoha, Kashiwa, Chiba 277-8568, Japan}
\footnotetext[\hokkaido]{Department of Cosmosciences, Hokkaido University, Kita-ku, Sapporo, Hokkaido 060-0810, Japan}
\footnotetext[\oklahoma]{H. L. Dodge Department of Physics \& Astronomy, University of Oklahoma, 440 W Brooks St Norman, OK 73019, USA}
\footnotetext[\tohoku]{Astronomical Institute, Tohoku University, Aoba-ku, Sendai, Miyagi 980-8578, Japan}

\title{Subaru/HiCIAO $HK_{\rm s}$ imaging of LkH$\alpha$ 330 - multi-band detection of the gap and spiral-like structures}

\begin{abstract}
We present $H$- and $K_{\rm s}$-bands observations of the LkH$\alpha$ 330 disk with a multi-band detection of the large gap and spiral-like structures. 
The morphology of the outer disk ($r\sim$$0\farcs3$) at PA=0--45$^\circ$ and PA=180--290$^\circ$ are likely density wave-induced spirals and comparison between our observational results and simulations suggests a planet formation.
We have also investigated the azimuthal profiles at the ring and the outer-disk regions as well as radial profiles in the directions of the spiral-like structures and semi-major axis.
Azimuthal analysis shows a large variety in wavelength and implies that the disk has non-axisymmetric dust distributions.
The radial profiles in the major-axis direction (PA=$271^\circ$) suggest that the outer region ($r\geq0\farcs25$) may be influenced by shadows of the inner region of the disk.
The spiral-like directions (PA=10$^\circ$ and 230$^\circ$) show different radial profiles, which suggests that the surfaces of the spiral-like structures are highly flared and/or have different dust properties. 
Finally, a color-map of the disk shows a lack of an outer eastern region in the $H$-band disk, which 
may hint the presence of an inner object that casts a directional shadow onto the disk.
%The morphology of the outer disk ($r\sim$$0\farcs3$) at PA=0--45$^\circ$ and PA=180--290$^\circ$ is affected by a density wave-induced spiral despite the possibility of ”cavity-like” structures.
\end{abstract}

\section{Introduction} \label{sec: Introduction} 

A protoplanetary disk loses almost all of its mass after 
a few million years \citep[][]{Haisch2001,Currie2009,Cloutier2014,Ribas2015},
during which the disk can be perturbed by accretion, jets, photoevaporation, dust growth, and planet formation \citep[e.g.,][]{Crida2007,Armitage2011}. 
Previous theoretical simulations 
%based on spectroscopic information (たぶん余計; modified by MUto)
predicted gaps within disks, which is likely related to planet formation \citep[e.g.,][]{Marsh1992,Rice2003,Zhu2011,Zhu2012}.
In recent years, dozens of high-spatial resolution observations have revealed a diversity of shapes within disks such as gaps, rings, and spiral features \citep[e.g.,][]{Hashimoto2011,Muto2012,ALMA2015,Perez2016}. Radio observations allow investigation of the gas and dust distributions of disk and infrared (IR) observations provide scattered light information from the surface of disks. 
Particularly, possible planet-disk interactions for many of these disk shapes have been suggested \citep[][]{Zhu2015,Dong2017}, as well as other predictions such as dust sintering \citep{Okuzumi2016}.
However, the number of reported companion candidates within the disks is much smaller than the number of asymmetric disks \citep[][]{Quanz2013hd100546,Reggiani2014,Currie2015hd100546,Sallum2015,Reggiani2017}.
Therefore continuing explorations of disk and companions is important for the study of planet formation and disk evolution mechanisms.

%comment by Muto
%referee から指摘されている "a broad introduction to the transitional disk class and pre-transitional disk class" に対応する文章が欲しい。
%Espaillat et al. の PPVI review も引いておけばどうでしょう？
% http://adsabs.harvard.edu/abs/2014prpl.conf..497E
% また、referee report は、「transitional disk だと思う」と書いているが、そこは改訂が必要。LkHa 330 は pre-transitional disk (中心にinner disk あり) ですよね。
%Added by Muto
A class of disks having an inner cavity, called transitional disks, is of particular interest in studying the disk evolution and dissipation.
Some of the transitional disks are expected to harbor a small inner disk that is optically thick in optical/near-IR around a central star and are called pre-transitional disks \citep[][]{Espaillat2010,Espaillat2014}. 
%Some of the transitional disks are expected to harbor an small inner disk around a central star and this inner disk is optically-thick in optical/near-IR. These disks are called pre-transitional disks 
In this paper, we report the result of near infrared scattered light imaging observations of LkH$\alpha$ 330.
LkH$\alpha$ 330 is a young stellar object (YSO) in the Perseus association ((RA,DEC) =(03 45 48.28, $+$32 24 11.9)). 
This system exhibits the spectral feature of a pre-transitional disk \citep{Espaillat2014} and \cite{Brown2007} suggested an inner disk by showing polycyclic aromatic hydro-carbon features. 
%A pre-transitional disk system has an optically-thick inner disk in the gap \citep[][]{Espaillat2010}.
Furthermore, mm and sub-mm observations have reported that the disk has a large ($\sim$$0\farcs16$--$0\farcs27$) gap \citep[][]{Brown2008,Andrews2011,Isella2013} and IR observations have suggested spirals \citep[][]{Akiyama2016}. 
These studies suggest grain growth and possibly unseen planets within the disk that may cause the large gap and spirals.
To follow-up the earlier studies on this intriguing system we conducted $H$- and $K_{\rm s}$-bands observations of LkH$\alpha$ 330 as a part of the Strategic Explorations of Exoplanets and Disks with Subaru \citep[SEEDS;][]{Tamura2009} 
project. 

In this paper we describe our observations and data reduction in Section \ref{sec: Observations and Data Reduction} and present our results and data analyses are shown in Section \ref{sec: Results and Data Analysis}. In Section \ref{sec: Discussions} we discuss the detected features in the disk and summarize our results.

\section{Observations and Data Reduction} \label{sec: Observations and Data Reduction}
The observations and data for LkH$\alpha$ 330, which are analyzed in this study, have been reported previously in \cite{Uyama2017}.

We made $H$-band ($\sim$1.6 $\mu$m) observations of LkH$\alpha$ 330 in 2014 October and $K_{\rm s}$-band ($\sim$2.2 $\mu$m) observations in 2015 January with Subaru/HiCIAO \citep{Suzuki2010} combined with classical AO system AO 188 \citep[][]{Hayano2008}. A typical FWHM of each unsaturated point-spread function (PSF) is $\sim$80 mas in the $H$-band and $\sim$70 mas in the $K_{\rm s}$-band.
The data of this study were obtained without a coronagraph for both the $H$-
and $K_{\rm s}$-band observations.
Note that \cite{Akiyama2016} also used
Subaru/HiCIAO to observe this system in 2011 December but adopted a $0\farcs4$ coronagraph mask, which prevents exploration of the gap region of the disk.
The data sets were taken by combining polarization differential imaging (PDI) to investigate faint disk structures and angular differential imaging (ADI) to detect substellar companions around the central star.
Detailed observation logs are shown in the SEEDS/YSO comprehensive report of
\cite{Uyama2017}, which applied only 
ADI reduction \citep[][]{Marois2006,Lafreniere2007} 
for companion explorations. This study focuses on the data analysis of PDI.
We used the ``quad-PDI" (qPDI) mode where two Wollaston prisms enable the acquisition of two ordinary and extraordinary rays on one frame simultaneously. Each field of view is $\sim$5\arcsec$\times\sim$5$\arcsec$ and the plate scale after distortion correction is 9.5 mas/pix.
After the first reduction of flat fielding, distortion correction, and image registration, all the polarimetric data sets were reduced in the same way as \cite{Hashimoto2011,Hashimoto2012} using IRAF.

\section{Results and Data Analysis} \label{sec: Results and Data Analysis}

% この段落は、一つの独立な小節 (Basic Parameters of LkHa 330 みたいなタイトル)として、セクション３の最初か、セクション２の最後あたりに書いておいた方が、話の流れが途切れないように思います。 (Comment by Muto)
%もし、この段落を小節にするなら、ここまでのセクション３の内容も、一つの小節にしておくと良いでしょう。例えば、「Polarized Intensity Image」 みたいなタイトルでどうかと思います。
\subsection{Basic Parameters of LkH$\alpha$ 330}

LkH$\alpha$ 330 was assumed to have 250 pc for a distance, 3 Myr old for an age, and GIII for a spectral type \citep[][]{Cohen1979,Enoch2006} in the previous studies introduced in Section \ref{sec: Introduction}. However, Gaia DR2 recently reported the distance to be $311\pm8$ pc \citep[][]{Gaia2016,GaiaDR2-2018}. Therefore we adopt the GAIA-measured distance in our discussion.  
\cite{Herczeg2014} estimated a spectral type of LkH$\alpha$ 330 to be F7.0 by analyzing its optical spectra with an assumed distance of 315 pc.
%By assuming that a GAIA-based spectral type does not largely vary from \cite{Herczeg2014} 
We converted the spectral type of F7.0 to the effective temperature using the relationship between a spectral type and effective temperature
in \cite{Pecaut2013}.
We used B and V-band magnitudes \citep{Mermilliod1987} to determine extinction-corrected V-band magnitude, which was converted to a LkHa 330's bolometric luminosity based on \cite{Pecaut2013}.
These effective temperature and luminosity described above were compared to the Pisa pre-main sequence evolution tracks \citep[][]{Tognelli2011}.
Finally, we estimate a mass and age of LkH$\alpha$ 330 to be $2.8\pm0.2 M_{\odot}$ and $2.5\pm0.7$ Myr, respectively. 
%This object was assumed 250 pc for a distance, 3 Myr old for an age, and GIII spectral type \citep[][]{Cohen1979,Enoch2006} in the previous studies above. However, Gaia DR2 reported the distance to be $\sim$310 pc \citep[][]{Gaia2016,GaiaDR2-2018}. Therefore we adopt 310 pc in our discussion.
%and calculated its luminosity to be $\log{L/L_{\rm{\odot}}} = 1.21$ with its assumed distance of 315 pc.   
%\cite{Herczeg2014} assumed 315 pc for its distance and estimated the spectral type at F7.0 by analyzing optical spectra.
%We corrected this luminosity to account for the difference of the Gaia-measured distance and the one in \cite{Herczeg2014}.
%and estimated the LkH$\alpha$ 330's mass and age with the Pisa pre-main sequence evolution tracks \citep[][]{Tognelli2011}.
%Note that we did not fully model the LkH$\alpha$ 330's spectral energy distribution (SED).

\begin{figure*}
\begin{tabular}{c}
\begin{minipage}{0.5\hsize}
 \centering
 \includegraphics[scale=0.6]{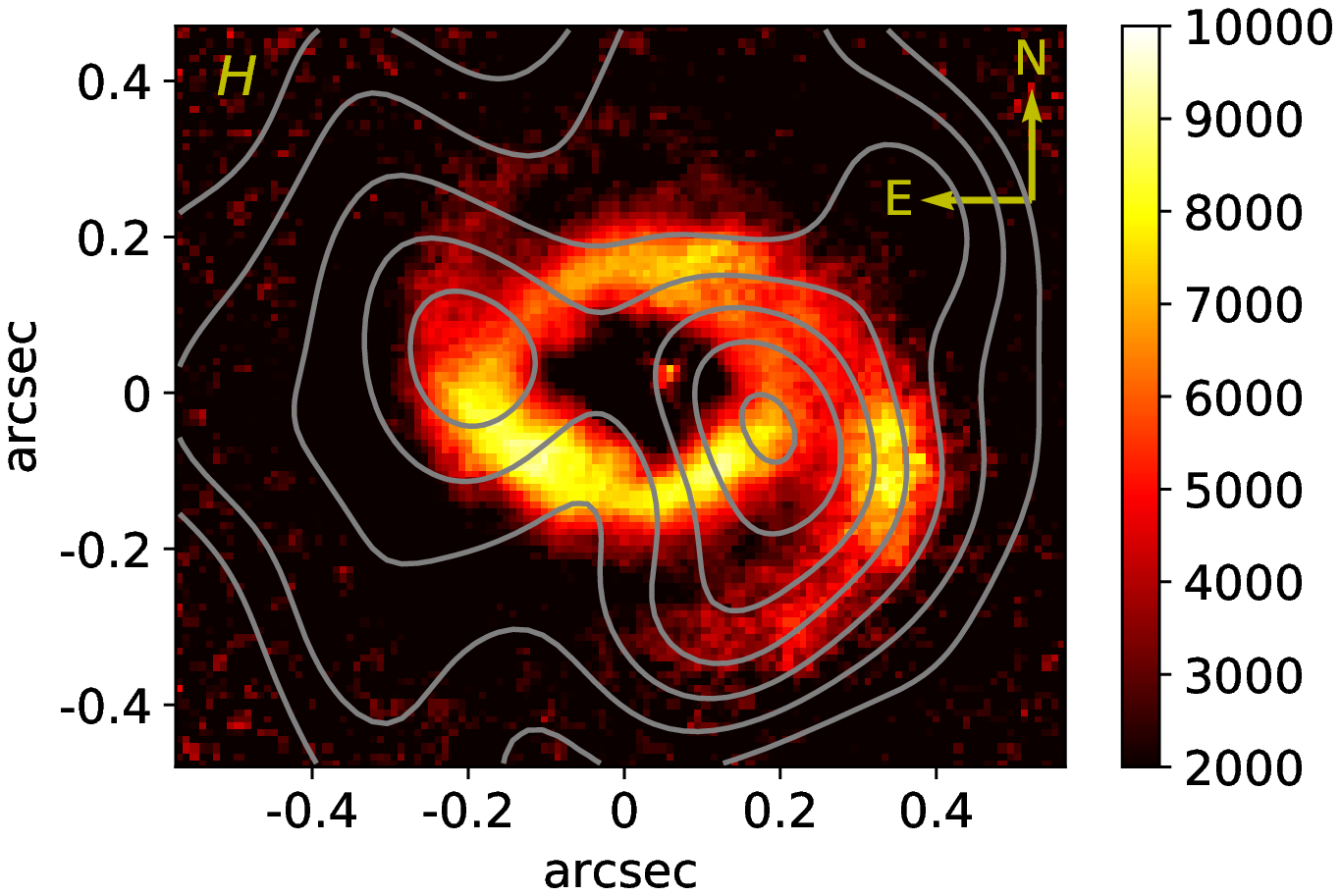}
\end{minipage}
\begin{minipage}{0.5\hsize}
 \centering
 \includegraphics[scale=0.6]{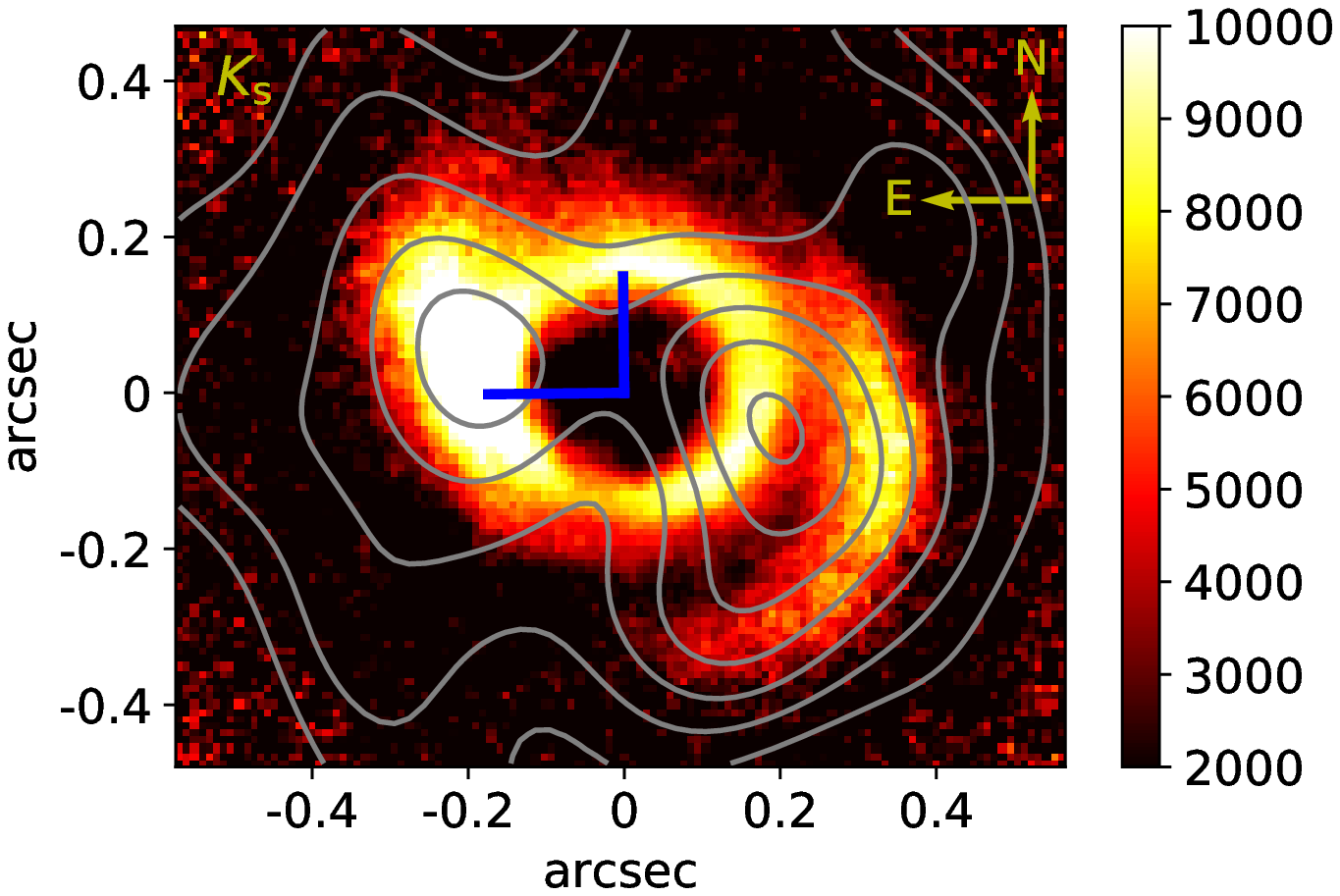}
\end{minipage}
\end{tabular}
\caption{Subaru/HiCIAO $r^2$-scaled PI image of LkH$\alpha$ 330 in the $H$- (left) and $K_{\rm s}$-bands (right). The scale bar is arbitrary. The SMA ($\lambda$=0.87 mm) observation result is overlaid on both images as gray contours. The levels are the same as those shown in \cite{Akiyama2016}. Perpendicular blue lines in the $K_{\rm s}$ image represent semi-major and semi-minor axes. North is up and east is left. }
\label{PI images}
\end{figure*}

\begin{figure*}
\begin{tabular}{cc}
\begin{minipage}{0.5\hsize}
 \centering
 \includegraphics[scale=0.4]{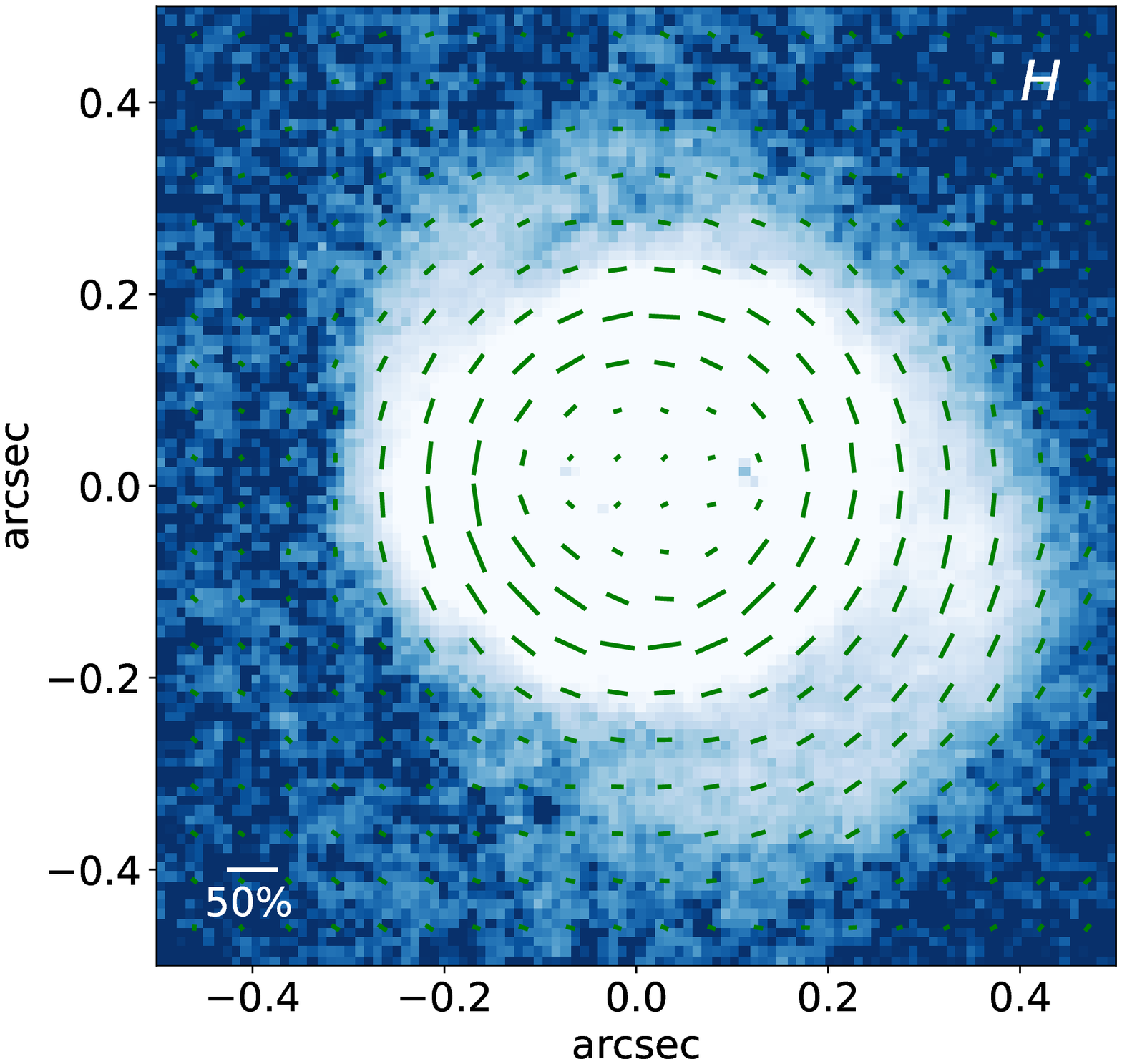}
\end{minipage}
\begin{minipage}{0.5\hsize}
 \centering
 \includegraphics[scale=0.4]{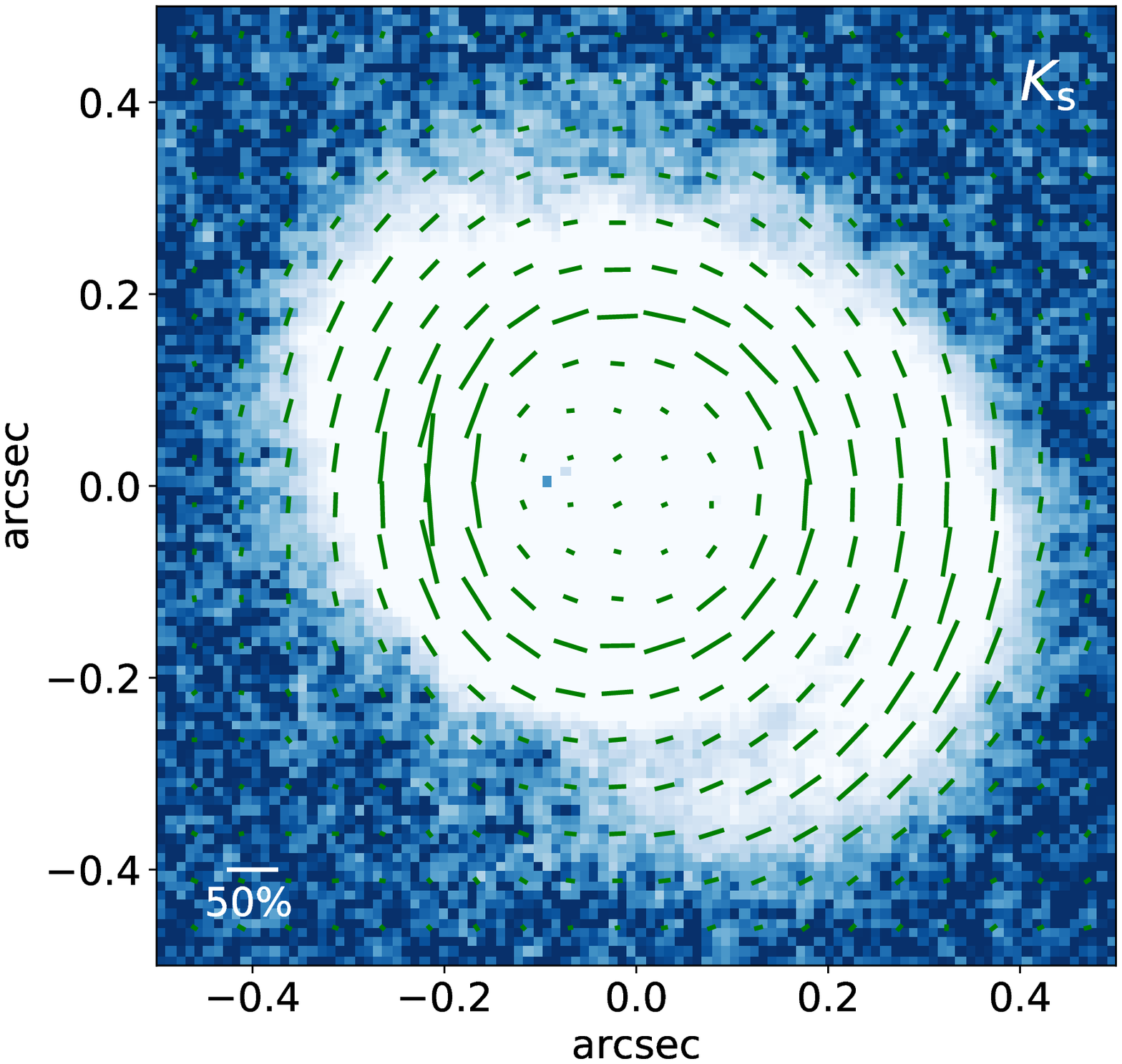}
\end{minipage}
\end{tabular}
\caption{$H$ and $K_{\rm s}$ Polarization vector maps superimposed on the PI maps.}
\label{pol vector}
\end{figure*}

\begin{figure}
 \centering
 \includegraphics[scale=0.55]{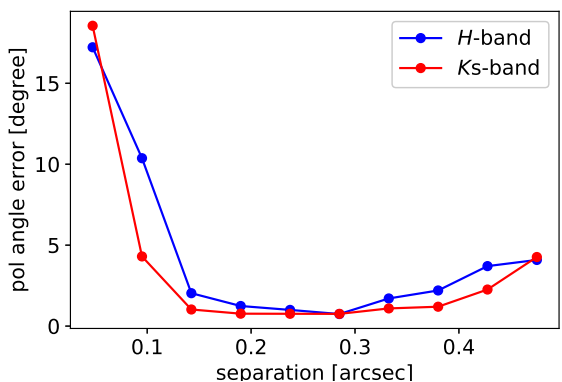}
 \caption{Polarization position angle errors as a radial profile in each image.}
 \label{pol vector error}
\end{figure}

\subsection{Polarized Intensity Image}

We detected a large gap and spiral-like features with signal-to-noise (S/N) ratios of $>$30 at the peak of the ring and $>$20 at the peak of the spiral regions. 
Figure \ref{PI images} shows $H$ and $K_{\rm s}$ $r^2$-scaled polarization intensity (PI) images. 
The PI parameter is dependent on $r^{-2}$ and we scaled this parameter by multiplying PI image by $r^2$ so that we can more properly see the structures. 
We note that this treatment corresponds to neglecting the disk flaring when calculating scattering angle, and may be problematic when discussing the disk structures quantitatively \citep{Stolker2016}.
%Modified by Muto
In this paper, we present some characteristic features on radial and azimuthal profiles and disk asymmetry of the disk, but keep the discussions on them at more or less qualitative levels.
%The original images are presented in the Appendix in a form of $U_{\rm r}$ and $Q_{\rm r}$. 
%We align the images on two color scales to make the gap and spirals clear. 
The contours taken from the Submillimeter Array (SMA) observation ($\lambda$=0.87 mm) discussed in \cite{Akiyama2016} are superimposed on the figures. 

% modified by Muto; 誤差計算のところで、段落分けたほうが良い（図１に、誤差が載っていないので。）
%When calculating 
To estimate the S/N ratio, we first measured the polarization vector as seen in Figure \ref{pol vector} and defined $\theta_{\rm err}$ as the 
difference of the angle between the polarization vector and the vector normal to the position vector measured from the central star.
%from the normal vector at delta-x and delta-y from the central star, which is briefly explained in Appendix. 
We then calculated radial profiles (see Figure \ref{pol vector error}) and converted the polarization angle error into a polarization error using an equation (6) in \cite{Kwon2016}.
In this estimate of the error, 
%In this calculation 
we assume that the "real" polarization vectors are centro-symmetric around the central star.
The observed polarization pattern is indeed very close to centro-symmetric (see Figure \ref{pol vector}) and the 
%a centro-symmetric pattern constitutes almost all the polarization as shown 
%in Figure \ref{pol vector} and 
non-azimuthal polarization such as T Cha \citep[][]{Pohl2017} is probably negligible.
We define noise as the standard deviation at given annular areas like ADI contrast limit \citep[see Section \ref{sec: ADI} and][]{Uyama2017}.
% 以下の節の言いたいことは何でしょう？(comment by Muto)
%but note that there can be seen azimuthal difference by a factor of $\sim$3 in Figure \ref{angle error map}. 

\subsection{Gap Region} \label{sec: Gap Region}
We traced the inner wall (hereafter we call this the ``ring") region in the $r^2$-scaled PI images and then fitted the peak profiles with an elliptic equation $(\frac{x-x_{\rm cen}}{a})^2+(\frac{y-y_{\rm cen}}{b})^2=1$, in which $a$ and $b$ are the semi-major axis and semi-minor axis. Table \ref{disk parameter} compares the cavity radius, position angle of the semi-major axis, and inclination from previous studies and this work. 
%(Modified by Muto)
In the calculation we used
the nonlinear least-squares (NLLS) Marquardt-Levenberg algorithm implemented into \verb+gnuplot+.  The error bars represent
%an implementation of the nonlinear least-squares (NLLS) Marquardt-Levenberg algorithm with gnuplot and error bar represents 
1$\sigma$ asymptotic standard error.
We note that previous studies used the midplane of the disk for measuring the gap while we used the surface brightness of the disk.
Observations of transitional disks have systematically revealed bigger cavity sizes in the mm continuum than in scattered light \citep[][]{Dong2012}, e.g., PDS 70 \citep{Hashimoto2012,Hashimoto2015} and 2MASS J1604 \citep[][]{Mayama2012,Dong2017J1604}. This phenomenon may be explained as being due to mm-sized dust being filtered out at the cavity edge due to gas-dust coupling effect \citep[e.g.,][]{Rice2006, Zhu2012, Dong2015gap}. 
Here we report that LkH$\alpha$ 330 is another example of this class of object. \cite{Andrews2011} modeled mm continuum emission of the disk and concluded that the cavity size is 84 au \citep[$0\farcs27$; see also][]{Isella2013}. The NIR cavity size seen by Subaru is only $\sim$54 au, much smaller than the mm cavity size.

%Our data reduction found a possible non-axisymmetric structure at $\rho\sim$$0\farcs1$ and position angle (PA) $\sim$160$^\circ$ in both images (see Appendix). An independent aperture masking observation reported a companion candidate at a similar separation but at another position angle \citep[$212\fdg9$;][]{Willson2016}. Since the S/N ratios of this structure in both PI images are less than 5, our data cannot be compared with this previous result but is a good motivation for follow-up observations in the gap region. 

\begin{table}  \centering
  \caption{Adopted parameters of the disk}
  \begin{tabular}{ccc}\\ \hline\hline
  Parameters & Previous study & This study \\ \hline
  semi-major axis [au]$^a$ & 50$^b$--84$^c$ & 54$\pm$1  \\
  %semi-minor axis [au] & $\dots$ & 34$\pm$2 \\
  position angle [degree] & 80$^b$ & 91$\pm$2 \\
  inclination [degree] & 35$^b$ & 31$\pm$3 \\ \hline
  \end{tabular}
  \label{disk parameter}
  \tablecomments{a: Angular separations do not change and thus we scaled the size reported from previous studies by the distance of 310 pc. b: \cite{Brown2008}, c: \cite{Andrews2011}}
\end{table}

\subsection{Spiral-like Region}
\begin{figure*}
\begin{tabular}{cc}
\begin{minipage}{0.5\hsize}
 \centering
 \includegraphics[scale=0.7]{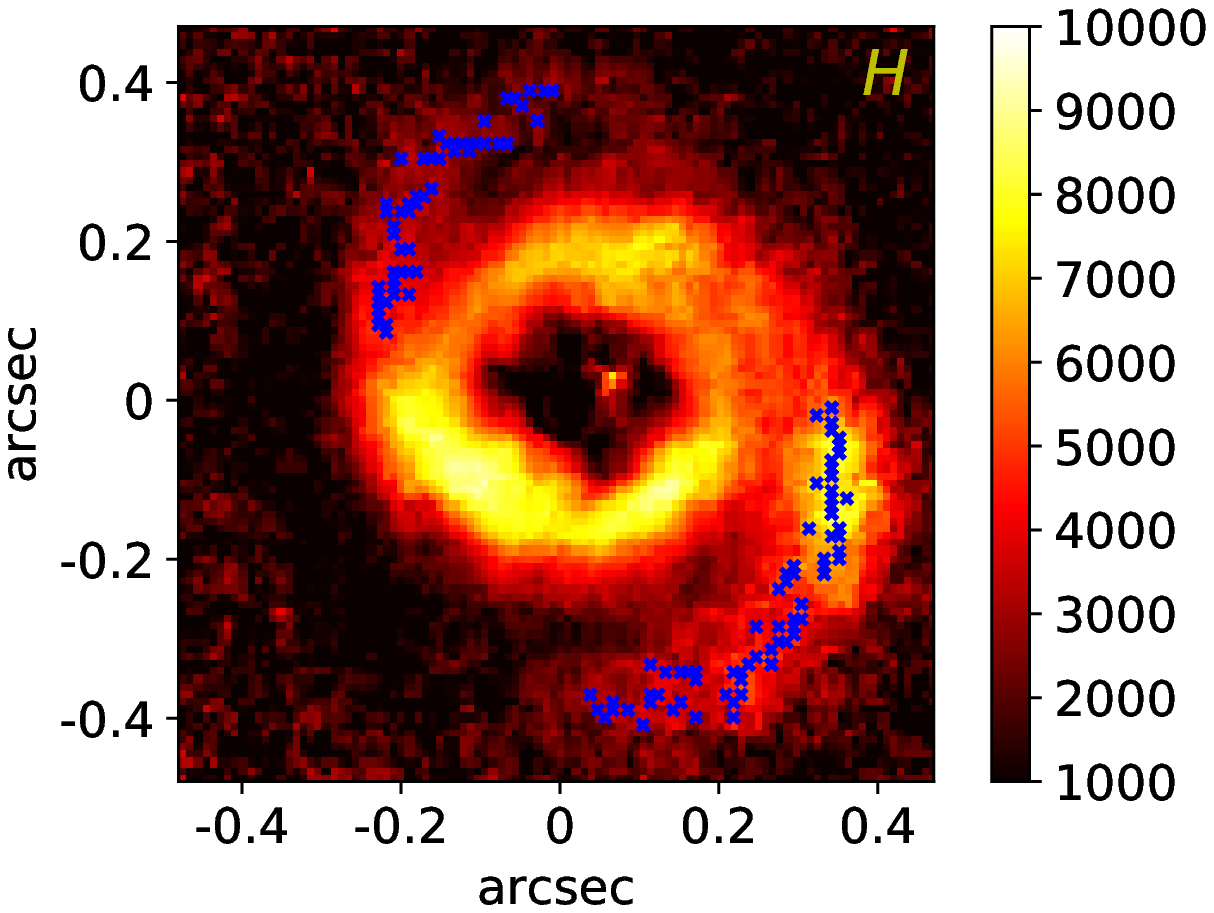}
\end{minipage}
\begin{minipage}{0.5\hsize}
 \centering
 \includegraphics[scale=0.7]{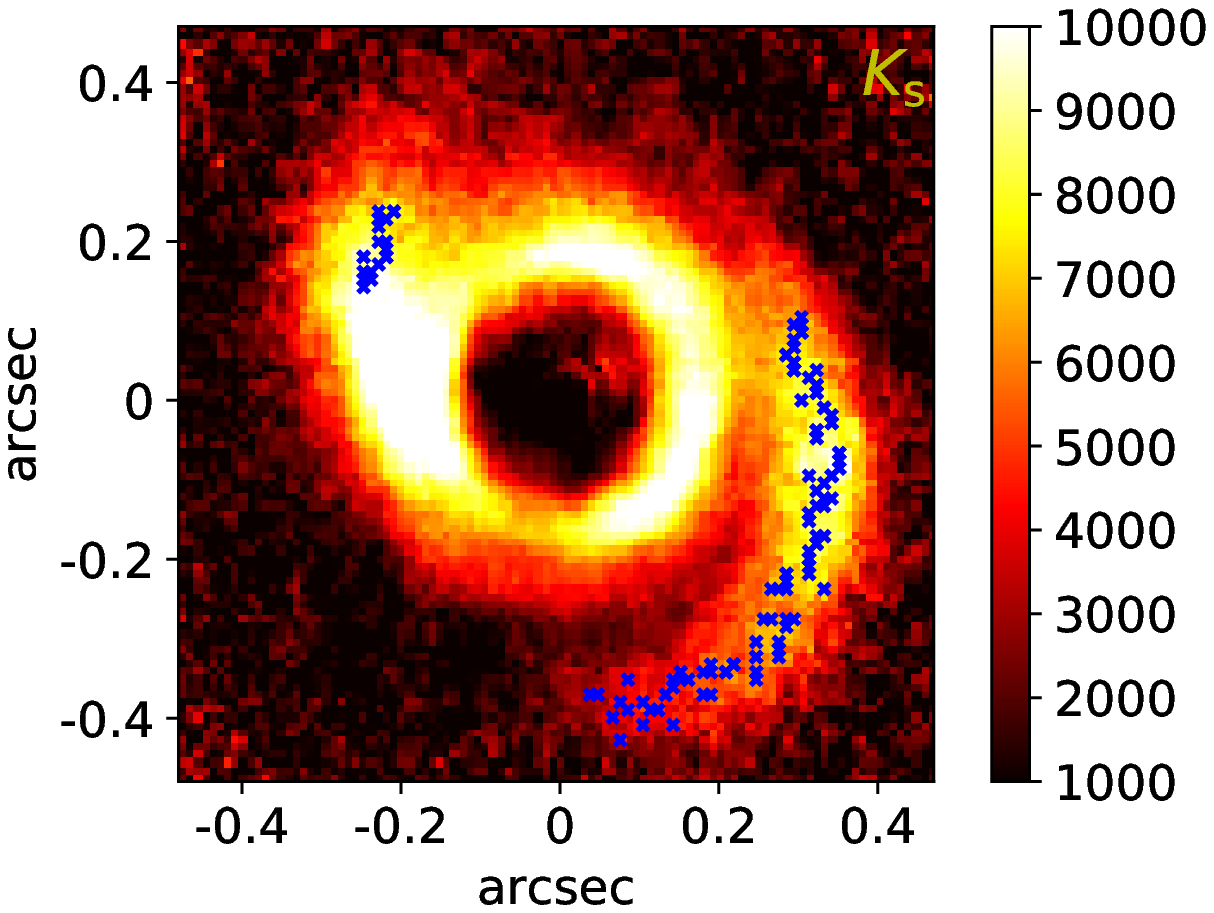}
\end{minipage}
\end{tabular}
\caption{Deprojected and $r^2$-scaled $H$- (left) and $K_{\rm s}$-band (right) PI images. We additionally plot traced peaks of the spiral-like structure in the south-west direction using blue crosses. Note that we have changed the scale bars from those in Figure \ref{PI images} in order to make the outer structures clearer.}
\label{spiral fit}
\end{figure*}

\begin{figure}
 \centering
 \includegraphics[scale=0.55]{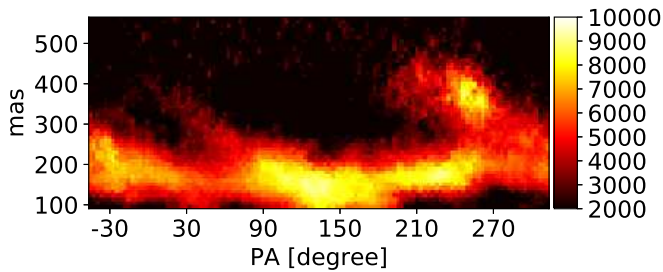}
 \caption{Polar-projected $H$-band PI image after deprojection and scaling by squared separations of $r^2$. The color scale is arbitrary. In order to make it easier to find outer asymmetric features, we started PA from $-45^{\circ}$. }
 \label{r-theta H}
\end{figure}

\begin{figure}
 \centering
 \includegraphics[scale=0.55]{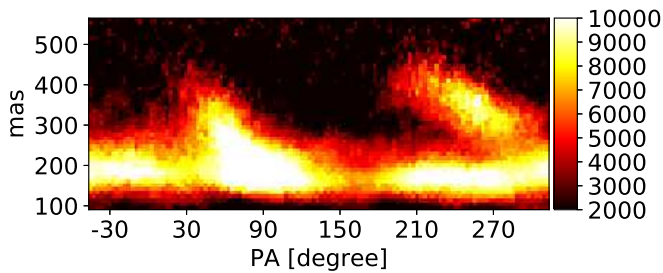}
 \caption{As Figure \ref{r-theta H} for the $K_{\rm s}$-band.}
 \label{r-theta K}
\end{figure}

As discussed in \cite{Akiyama2016} the two peaks of the SMA continuum are located at spiral-like features. Interestingly, the surface and midplane distributions at the south-west region are consistent, while the north-east region does not exhibit a similar distribution.
Figure \ref{spiral fit} shows deprojected and $r^2$-scaled PI images and figures \ref{r-theta H} and \ref{r-theta K} show polar-projected images taken from Figure \ref{spiral fit}.
We traced the ridge of the spiral-like structures, which is superimposed on the images (blue crosses). 
The morphology of the outer region implies that the disk's rotation is counter-clockwise. 
Note that we did not change the scattering angle when deprojecting the PI images but changed their inclinations to zero only for the purpose of tracing the peaks of the outer structures.

%We can now see clear deviations at two positions, one about a third from the left side and the other near the edge of the right side. The left and right deviations correspond to the north-east and south-west structures, respectively. 
We can now see clear deviations from the axisymmetric ring-like structure at $r$$\sim$200 mas. There are two spiral structures: one is launched at about PA=290$^\circ$ and the other is at PA=70$^\circ$.
We find that the $H$- and $K_{\rm s}$-band observations have different shapes of the outer asymmetric features. 
For the south-west non-axisymmetric features, they extend from PA$\sim$290$^\circ$ to 180$^\circ$ both in $H$- and $K_{\rm s}$-bands, and they appear like "spiral" features. 
For the north-east feature, however, the H-band feature extends from PA$\sim$70$^\circ$ to 0$^\circ$ and it does appear like a "spiral", while in the $K_{\rm s}$-band, the emission between PA$\sim$40$^\circ$ and 0$^\circ$ is missing. The appearance of the north-east feature in the $K_{\rm s}$-band may be described as "slightly inclined blob".
Anyway, the inclined spirals can have complex morphology \citep[e.g.,][]{Dong2016} and we discuss the possibility of the spiral in Section \ref{sec: Spiral-like Structures}.

We could trace the peaks of the south-west structure between PA=180$^\circ$--270$^\circ$ in the $H$-band and between PA=180$^\circ$--290$^\circ$ in the $K_{\rm s}$-band.
On the other hand, we could trace the peaks of the north-east structure between PA=0$^\circ$--70$^\circ$ in the $H$-band and between PA=40$^\circ$--60$^\circ$ in the $K_{\rm s}$-band.
We investigated angles between the roots of the spirals and the ring by using the $H$-band result for north-east structure and the $K_{\rm s}$-band result for south-west structure.
The pitch angles are determined to be $\sim$12$^\circ$ and $\sim$16$^\circ$ for the south-west and north-east structures, respectively. These values are similar to the SAO 206462 spirals \citep[][]{Muto2012}.
%We could not trace the peaks of the other spiral-like structure due to its ambiguous morphology.
%by fitting the PA=270$^\circ$--290$^\circ$ crosses in the $K_{\rm s}$-band with a line and comparing the normal vector at $290^\circ$

However, we note a weak tendency from the ``spiral top" toward the other side of disk in Figures \ref{r-theta H} and \ref{r-theta K}, which possibly represents another mechanism.
A trailing spiral behaves as a monotonically increasing profile \citep[][]{Goldreich1979}. The spiral density wave is likely to be a trailing feature, and it may be difficult to explain the blob-like morphology that we see in the north-east in the Ks-band image.
Since previous SMA and CARMA observations did not report any specific features except for the central large gap, identifying these asymmetric outer structures requires follow-up observations with a high spatial resolution.

%In the $H$-band the north-east feature is possibly a “cavity”
%A trailing spiral behaves as a monotonically increasing profile \citep[][]{Goldreich1979}. The wave density theory cannot produce a spiral where both trailing and leading spirals appear simultaneously. However, another possibility is that it is a planet-induced spiral. The disk is inclined and planet-induced spirals can have complex morphology \citep[e.g.,][]{Dong2016}. 
%The south-west feature looks like a spiral. %but is ambiguous. 

%In the $H$-band both of the outer features look like planet-induced spirals, while in the $K_{\rm s}$-band the north-east feature looks like a mere extended region of the outer disk, while the south-west feature is likely a spiral. The disk is inclined and planet-induced spirals can have complex morphology \citep[e.g.,][]{Dong2016} and we discuss the possibility of the spiral in Section \ref{sec: Origin of the Asymmetric Structures}.
%

%In Section \ref{sec: Origin of the Asymmetric Structures} we assume these structures to be spirals in order to discuss their origin but we leave open the possibility of other complicated structures.

% PA coverage: 180-270 for H, 180-290 for K


\subsection{Azimuthal Profiles} \label{sec: Azimuthal Profiles}
\subsubsection{Ring Region} \label{sec: Ring Region}

\begin{figure*}
 \centering
 \includegraphics[scale=0.48]{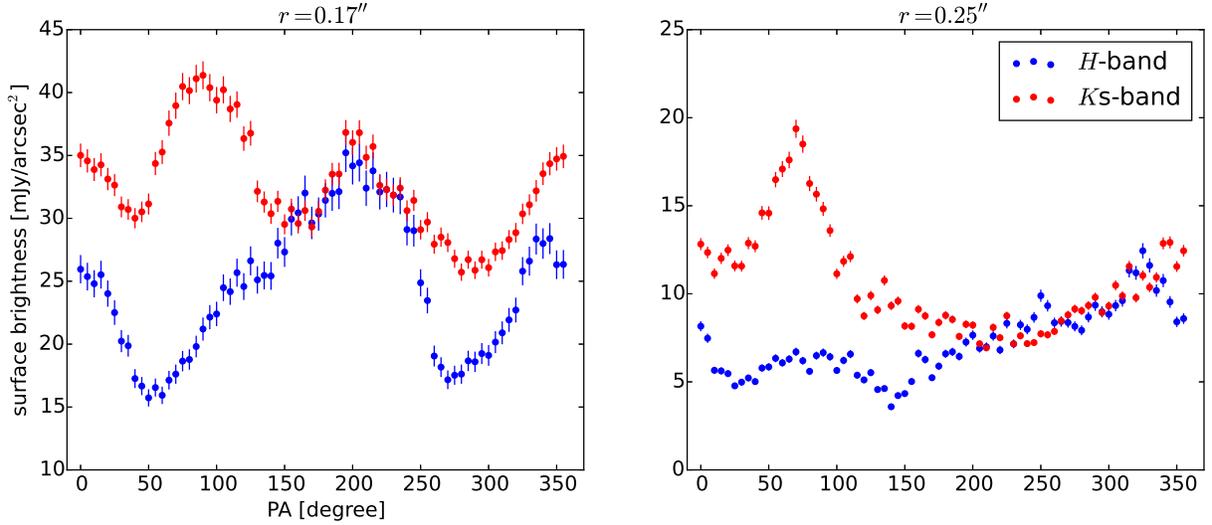}
 \caption{Azimuthal surface brightness profiles at $\sim$$0\farcs17$ (left) and $\sim$$0\farcs25$ (right).}
 \label{azimuthal profile of outer disk}
\end{figure*}

%Modified by Muto
We investigated the surface brightness profiles of the deprojected 
%disk 
image after averaging 
%each 
over 5$\times$5 pix so that we can expect to reduce 
the noise at the pixel scale.
%the influence of selection biases of the pixels. 
%1ピクセルでは少々差があるも、3×3とはほぼ差なし。全体的な形状の差は出なかった。
Figure \ref{azimuthal profile of outer disk} shows azimuthal profiles of the deprojected PI images at separations of $r=0\farcs17$ and $0\farcs25$.
We find that azimuthal profiles at the ring region ($r=0\farcs17$) 
show strong wavelength dependence. This is in contrast to the results of multi-band observations of the disk in other systems
%vary among wavelengths compared to other multi-band disk observations 
\citep[e.g.,][]{Bensity2017}. 
The south direction (PA$\sim$180$^\circ$) is the brightest in the $H$-band profile.
Considering that the minor axis of the disk is in the north-south direction (see Section \ref{sec: Gap Region}), this is probably due to the excess of forward scattering  
and therefore the southern side of the disk is likely to be the near side.
However, in the $K_{\rm s}$-band, the scattered light is the brightest in the eastern side, which is along the major axis of the disk. Such a large variation in the scattered light profiles in different bands might suggest that the dust distribution is not azimuthally symmetric. This issue will be further discussed in Section \ref{sec: Scattering Properties}.
%In the $K_{\rm s}$-band, however, the east direction (PA$\sim$90$^\circ$) is the brightest, which may not support forward scattering idea. Non-azimuthal distribution of dust particles may need to be taken into account to explain the difference of the $H$-band and $K_{\rm s}$-band images.  This issue will be further discussed in Section \ref{sec: Scattering Properties}.
%Four dot-dashed curves in each graph are expected phase functions with simple scattering properties. 
%%These difference can be induced by complicated dust distributions and details are

\subsubsection{Outer Region} \label{sec: Outer Region}
Our $H$-band profile at $r=0\farcs25$ is different from \cite{Akiyama2016}, particularly at the north region. The previous observation used the coronagraph mask and did not explore the central region, while our observation could explore much inner region.
Our profile might be partially influenced by the asymmetric dust distribution of the ring. 
%コロナグラフの有無の差に言及
We also find that these profiles are quite different from the ring azimuthal profile, particularly the relative decline of the surface brightness at the forward scattering region.
The spiral-like features appear at $r\sim$$0\farcs3$ in Figures \ref{r-theta H} and \ref{r-theta K} but our data sets suggest that the $r\sim$$0\farcs25$ area likely belongs to the outer spiral-like region.

\subsection{Radial Profiles}

\begin{figure*}
 \centering
 \includegraphics[scale=0.44]{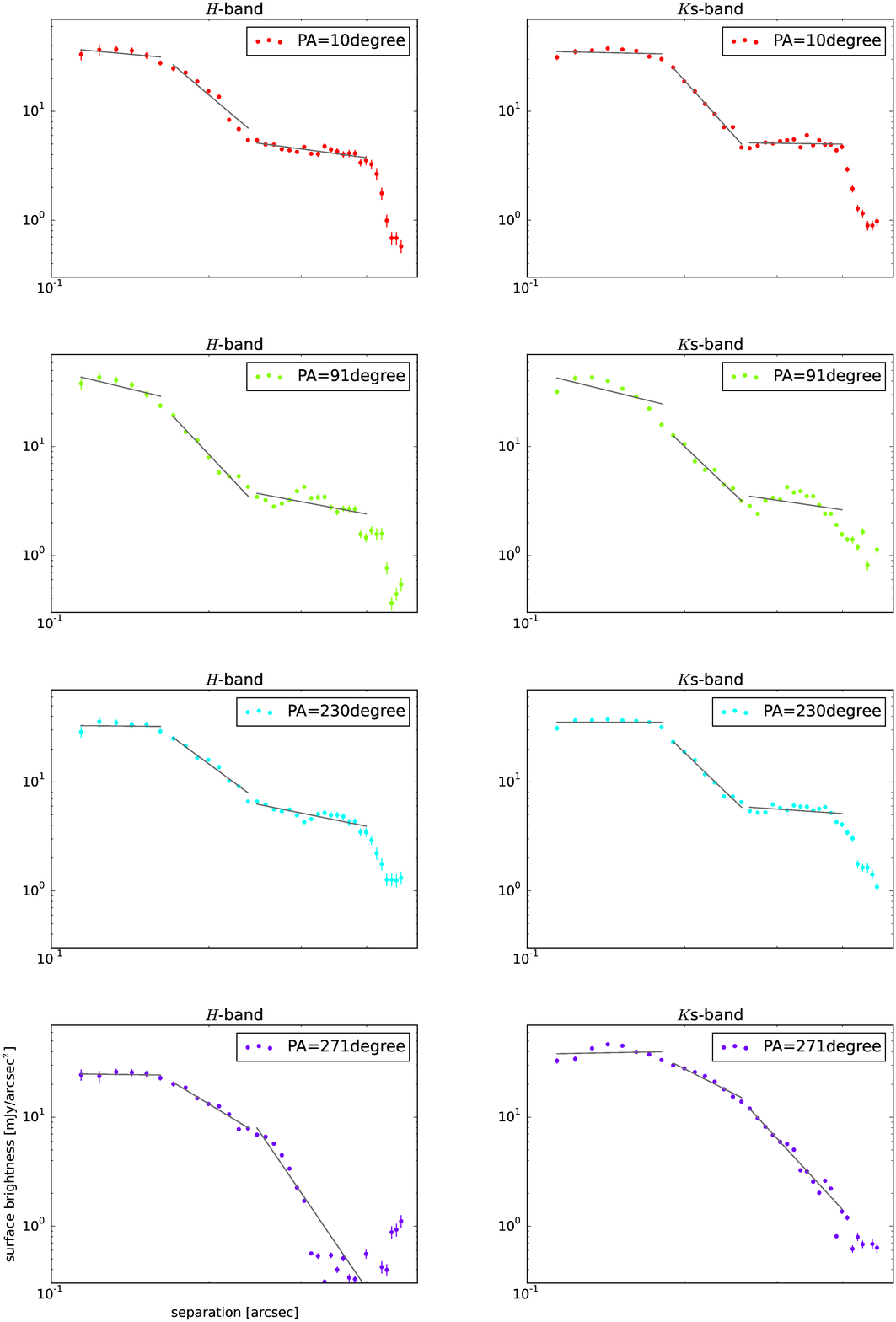}
 \caption{Radial profiles in the $H$- and $K_{\rm s}$-band data sets. The gray lines are power-law-fit results at three separations. The fitted results are shown in Table \ref{power-law fit}.}
 \label{radial profile}
\end{figure*}

Figure \ref{radial profile} shows radial profiles between $0\farcs15$ and $0\farcs45$ in both the $H$- and $K_{\rm s}$-band PI images. These profiles have azimuthal asymmetry.
The PAs 10$^\circ$ and 230$^\circ$ correspond to both spiral-like features. $91^\circ$ and $271^\circ$ correspond to the semi-major axis directions. 
We used the least squares method on these profiles and a power-law to investigate the surface structure of the disk. 
The fitted powers are listed in Table \ref{power-law fit}.
Except at the ring and spiral-like regions, the surface brightness is in proportion to the separation, to the power of no larger than 2, and is different from a flared disk's behavior.
An $r^{-3}$ profile can be explained with a flat disk \citep{Momose2015}, which can produce shadows due to surface structures and make the surface brightness profiles complex. 

Within $r < 0\farcs26$ the $H$ and $K_{\rm s}$ profiles are similar but have small difference. This difference may reflect a difference in the ring's scattering between the $H$- and $K_{\rm s}$-bands. In both bands the inner regions (H: $\leq 40$ au, K: $\leq 45$ au) behaves as highly flared disks, which creates shadows outward (H:50--75 au, K: 55--80 au). 
By combining these features we can assume that the $K_{\rm s}$-band ring extends more than the $H$-band ring.

At the outer region the fitted powers are smaller than $-2$ along PA=$271^\circ$ direction, which suggests that the disk's surface in the semi-major axis directions also behaves as a flat disk influenced by shadows. PA=$91^\circ$ profile is the semi-major axis direction but mixed with the spiral-like structure.
In PA=10$^\circ$ and 230$^\circ$ directions the profiles exhibit a more gradual change, with a steep decrease at separations greater than $r>0\farcs4$.
These profiles are consistent with flaring of the spiral-like structure and these region may have different dust properties. 

%Furthermore, the radial profiles have azimuthal asymmetry and thus we did not conduct synthetic modeling with the obtained radial and azimuthal profiles.

%Within $r < 0\farcs25$ the profiles in $H$ and $K_{\rm s}$ are quite different. This difference can reflect ring's scattering difference between $H$- and $K_{\rm s}$ bands. In $H$-band the inner region ($\leq 50$ au) behaves as a standard-flat disk and the mid region (between 50 and 60 au) does as a flat disk and/or being influenced by shadows. In $K_{\rm s}$ band the inner region ($\leq 45$ au) behaves as a highly flared disk, which makes longer shadows between 45 and 65 au. By combining these features we can assume that $K_{\rm s}$-band ring extends less than that in $H$-band but instead flares higher.

\begin{table*}  \centering
  \caption{Power-law fit of radial profiles}
  \begin{tabular}{ccccccc}\\ \hline\hline
  PA [degree] & \multicolumn{3}{c}{$H$} & \multicolumn{3}{c}{$K_{\rm s}$} \\ 
  & $0\farcs1 < r \leq 0\farcs15$ & $0\farcs15 \leq r \leq 0\farcs24$ & $0\farcs24 \leq r \leq 0\farcs40$ & $0\farcs1 < r \leq 0\farcs18$ & $0\farcs18 \leq r \leq 0\farcs26$ & $0\farcs26 \leq r \leq 0\farcs40$ \\
  \hline
 10&-0.44&-4.1&-0.65&-0.10&-5.4&-6.5$\times10^{-2}$\\
 %83&-0.90&-3.0&-7.2&-0.80&-3.7&-5.7 \\
 91&-1.2&-5.1&-0.91&-1.2&-4.6&-0.70 \\
 230&-5.0$\times10^{-2}$&-3.5&-1.0&7.8$\times10^{-3}$&-4.6&-0.33 \\
 %263&-0.88&-3.4&-3.6&0.57&-4.8&-3.7   \\ \hline
 271&-6.5$\times10^{-2}$&-2.9&-7.0&9.0$\times10^{-2}$&-2.4&-5.2 \\ \hline
  \end{tabular}
  \label{power-law fit}
\end{table*}

\subsection{Angular Differential Imaging} \label{sec: ADI}
\begin{figure}
 \centering
 \includegraphics[scale=0.4]{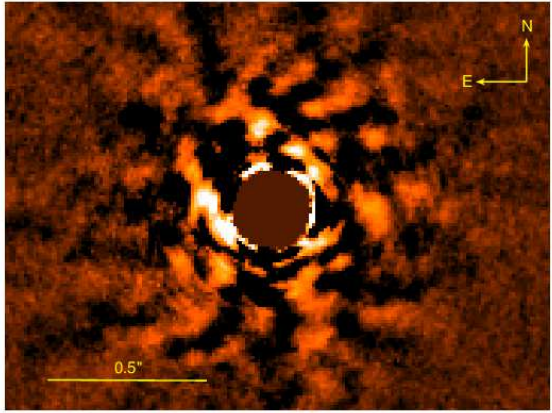}
 \caption{ADI-reduced image in the $K_{\rm s}$-band. North is up and the central star is masked by the algorithm.}
 \label{ADI result}
\end{figure}

As companion exploration, \cite{Uyama2017} conducted an ADI reduction of all the LkH$\alpha$ 330 data sets and could not find any companion candidates around the central star. Figure \ref{ADI result} shows our ADI-reduced image of the $K_{\rm s}$-band observation. We performed ADI-LOCI reduction \citep{Lafreniere2007} and the algorithm automatically masked a $\sim$$0\farcs15$ region from the center.
Our ADI-LOCI pipeline did not work properly 
for the nearly face-on disk, resulting in artificial residual pattern as seen in Figure \ref{ADI result}. Therefore, one cannot discuss the morphology of the disk in the ADI-LOCI reduced image. Nonetheless, the image can be used
to constrain the flux from a possible point-like source. Some signals remain around the central star but their S/N ratios of them are less than 5 and therefore we regard them as being residual from the PSF subtraction.

\begin{figure}
 \centering
 \includegraphics[scale=0.6]{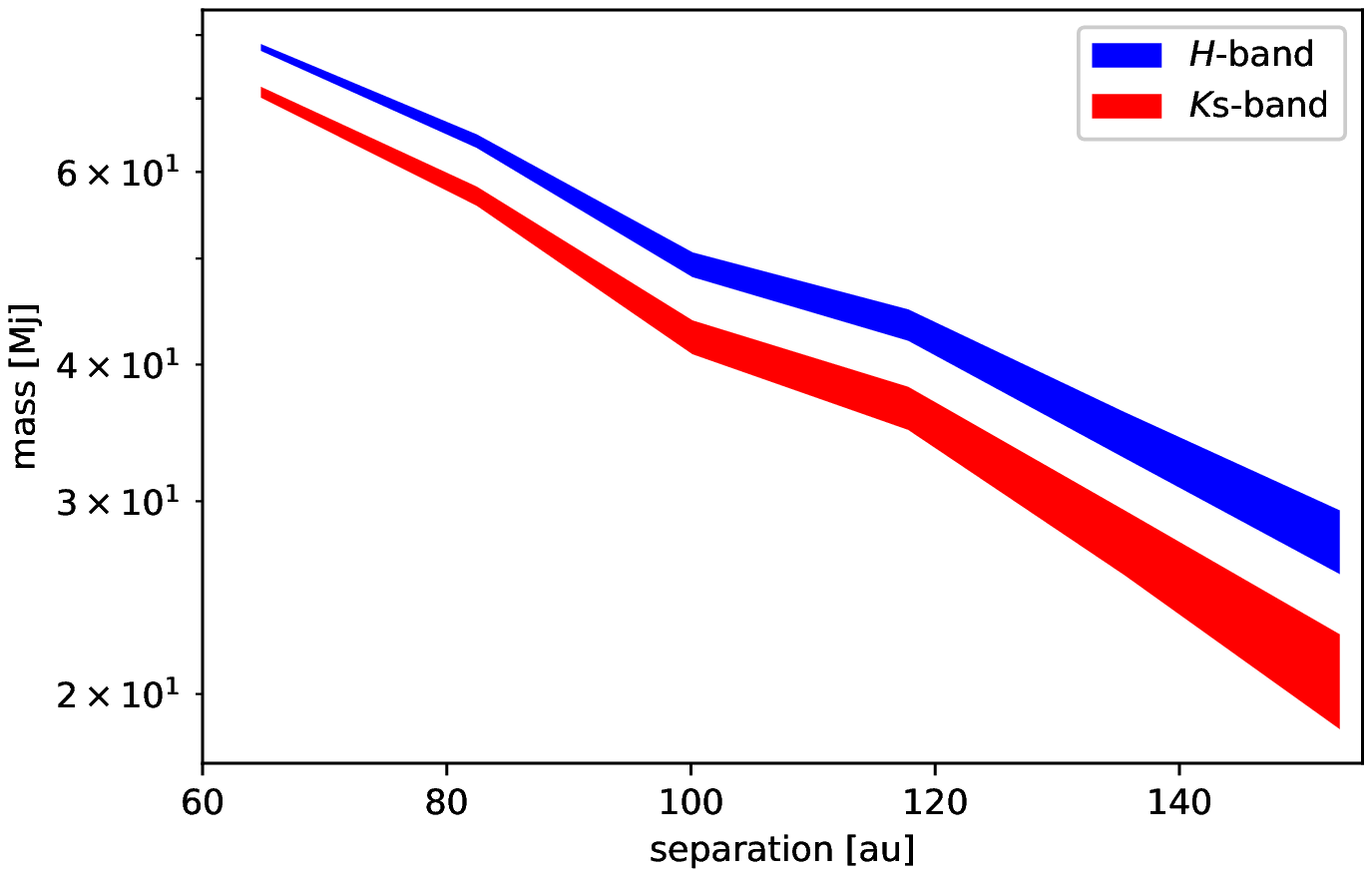}
 \caption{Mass limits of the ADI-reduced images. The vertical axis is mass in $M_{\rm J}$ unit and the horizontal axis is the projected separation. We converted the contrast limit into mass units by the BT-Settl model assuming that possible companions can clear the gas and dust locally and that the conversion can ignore extinction from the disk.}
 \label{detection limit}
\end{figure}

The contrast limits have already been described in \cite{Uyama2017}, and
were converted into mass limits based on the COND03 model \citep{Baraffe2003}.
%Modified by Muto
Figure \ref{detection limit} shows the mass limit of our observations.  
We have set
%and could set 
constraints on the mass of potential companions in the disk down to $\sim$20 $M_{\rm J}$.
%object within the disk ($M_{comp}$/$M_{\star}$ $\gtrsim$ 0.01).
The shadows correspond to errors of our age estimation. The conversion was based on the ``hot-start" model \citep[BT-Settl;][]{Allard2011}.

Besides age, planet mass limits could also be uncertain due to assumed planet luminosity evolution models.   
%These in turn depend on the initial conditions for the planet formation process and, more broadly, the mode of formation.  
``Hot-start" evolutionary models such as those we adopt are often associated with disk instability formation. 
``Cold-start" models \citep[][]{Marley2007} attempt to model planet formation by core accretion and yield lower initial entropies and luminosities and thus higher masses for a given contrast limit \citep[though see][]{Berardo2017}.
However, demographics suggest that companions with these masses/mass ratios and separations detectable from our data are likely not planets formed by core accretion \citep{Brandt2014,Currie2011}. Thus, our mass limits are likely to probe only companions formed like binary stars or by disk instability.   
%``Hot start'' evolutionary models have a high entropy initial state and luminosities and are often been associated with disk instability formation. ``Cold start'' models \citep[][]{Marley2007} attempt to model planet formation by core accretion and yield lower initial entropies and luminosities.  ``Warm-start" models \citep[][]{Spiegel2012} have an intermediate luminosity evolution. While demographics suggest that that both formation mechanisms are plausible for known directly-imaged planets \citep{Brandt2014,Currie2011}, core accretion-formed planets may have a ``hot start"-like luminosity evolution \citep{Berardo2017}.
%the newest luminosity evolution models suggest there may be no for at least some directly-imaged planetary-mass bodies may form by core accretion and others by disk instability.  
%As many of high-contrast imaging observations mentioned, evolutionary models depend on initial conditions. Given that even using the hot-start model could not constrain planetary-mass planets, we do not use any other models. The "warm-start" \citep[][]{Spiegel2012} and "cold-start" \citep[][]{Marley2007} models assume planet formation with smaller initial entropy and radii. 

%We used 2.85 for Av \cite[][]{Herczeg2014} instead of 1.8 that was previously assumed in \cite{Uyama2017} because of the $\sim$20\%-increase of the distance. 
%However, our data reduction could not explore the gap region due to inner working angle and the distance of this system.

\section{Discussions} \label{sec: Discussions}
The disk around the LkH$\alpha$ 330 has complex morphology. In this section, we focus on disk features one by one and discuss the implications on the disk properties, which will help to synthetically model the disk and to investigate disk evolution mechanisms such as gap opening and spiral forming with unseen planets.

\subsection{Morphology} \label{sec: Origin of the Asymmetric Structures}
%citation変更、gap originの話追加(evernote)
\subsubsection{Gap}
Previous observations of LkH$\alpha$ 330 suggested planet formation within the gap \citep[e.g.,][]{Zhu2012,Isella2013}. Our ADI reduction could not fully explore the gap region and thus planet formation remains a plausible but unconfirmed scenario.

Grain growth \citep[e.g.,][]{Brinstiel2012} and disk wind \citep[e.g.,][]{Suzuki2010diskwind} are also possible mechanisms for opening the gap in the disk.
A spectral feature of LkHa 330 is an excess at the mm and sub-mm wavelength ranges \citep[][]{Brown2008,Hitchcock_prep}, which suggests the existence of larger dust and supports the possibility of grain growth.
Investigating disk wind will require follow-up observations of gas kinematics; the presence of disk wind is suggested if blue-shift components excel in the data.
Although photoevaporation can produce a gap within the disk \citep[e.g.,][]{Clarke2001,Goto2006,Owen2011}, the disk mass is much larger \citep[$M_{\rm disk}\sim0.01M_\star\geq0.02M_\odot$;][]{Andrews2011} than expected mass of the photoevaporation stage \citep[an order of 0.001$M_\odot$;][]{Alexander2006}.

%this mechanism is not likely to form such a large gap considering the very short timescale of gap opening by photoevaporation \citep{Alexander2006}.

\subsubsection{Spiral-like Structures} \label{sec: Spiral-like Structures}

The LkH$\alpha$ 330 system may be another disk, after SAO 206462 \citep{Muto2012}, MWC 758 \citep{Grady2013}, HD 100453 \citep{Wagner2015}, DZ Cha \citep{Canovas2018} (note that the inner disk in the AB Aur system, \citealt{Hashimoto2011}, has also been suggested to host two spiral arms, Figure 14 in \citealt{Dong2016}), that has been discovered to have a pair of nearly symmetric spiral arms in scattered light. 
Except for HD 100453, which has an M dwarf companion that is probably driving the arms \citep{Dong2016HD100453,Wagner2018}, the origin of the spiral arms in the other systems is under debate. 
Two mechanisms are proposed to explain the morphology of these near $m=2$ arms: gravitational instability \citep[e.g.,][]{Dong2015spiral}, and companion-disk interaction \citep[e.g.,][]{Dong2015spiral, Zhu2015}. As LkH$\alpha$ 330's disk mass is perhaps too low to trigger the gravitational instability \citep[$M_{\rm disk}\sim0.01M_\star$;][]{Andrews2011}, we consider companion-disk interactions to constitute a plausible scenario.

\begin{figure*}
\begin{tabular}{ccc}
\begin{minipage}{0.33\hsize}
 \centering
 \includegraphics[scale=0.44]{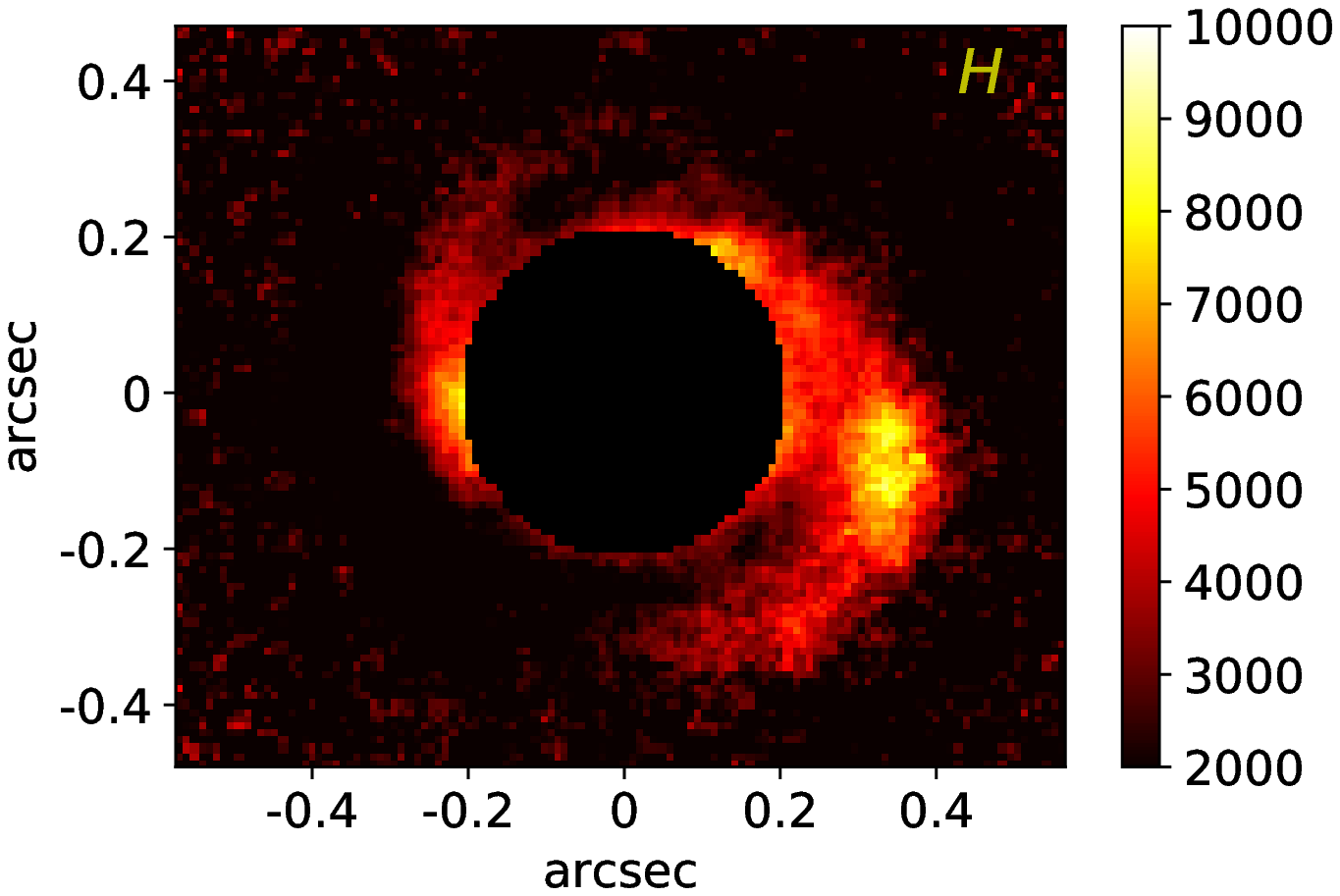}
\end{minipage}
\begin{minipage}{0.33\hsize}
 \centering
 \includegraphics[scale=0.44]{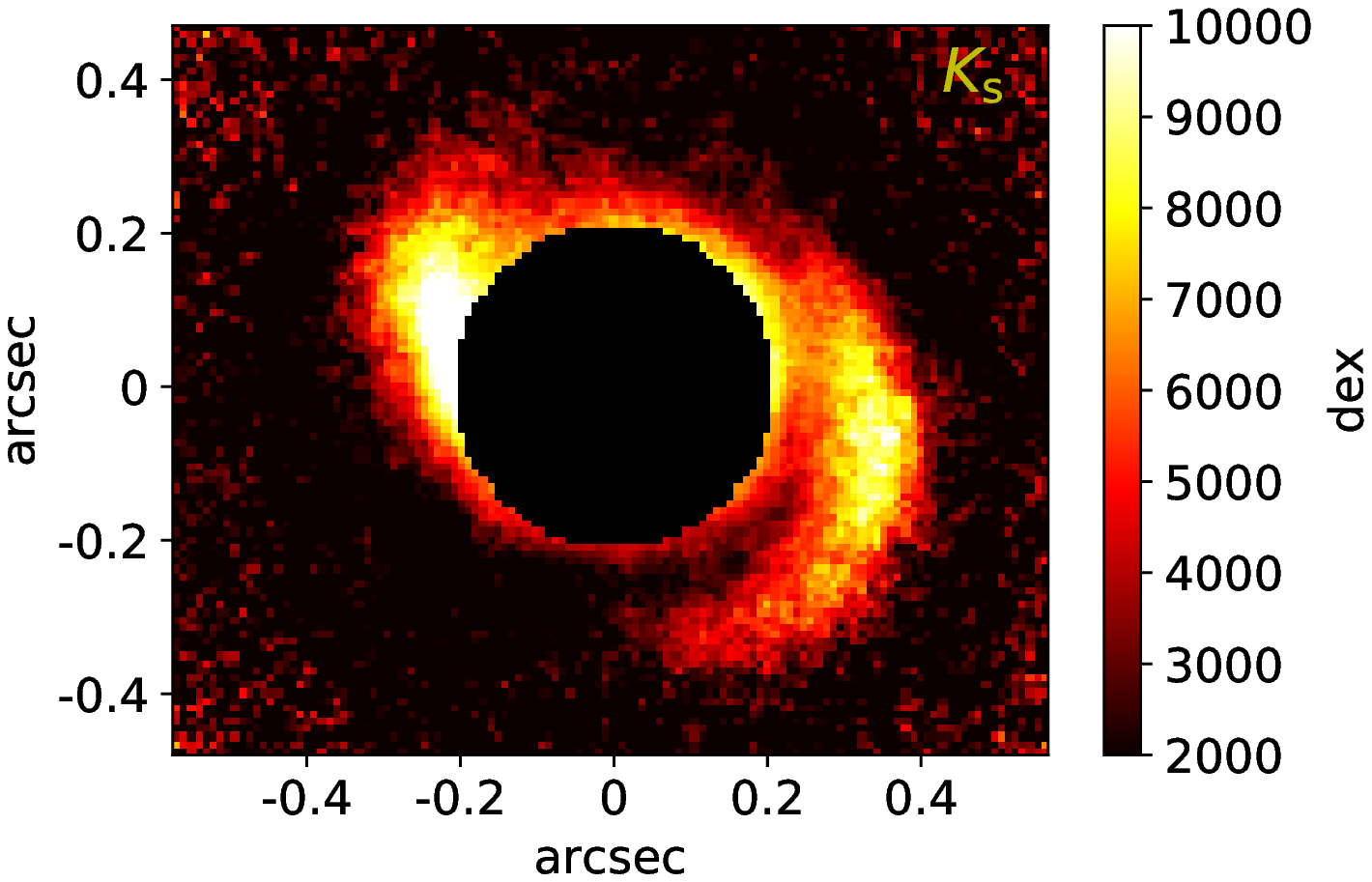}
\end{minipage}
\begin{minipage}{0.33\hsize}
 \centering
 \includegraphics[scale=0.43]{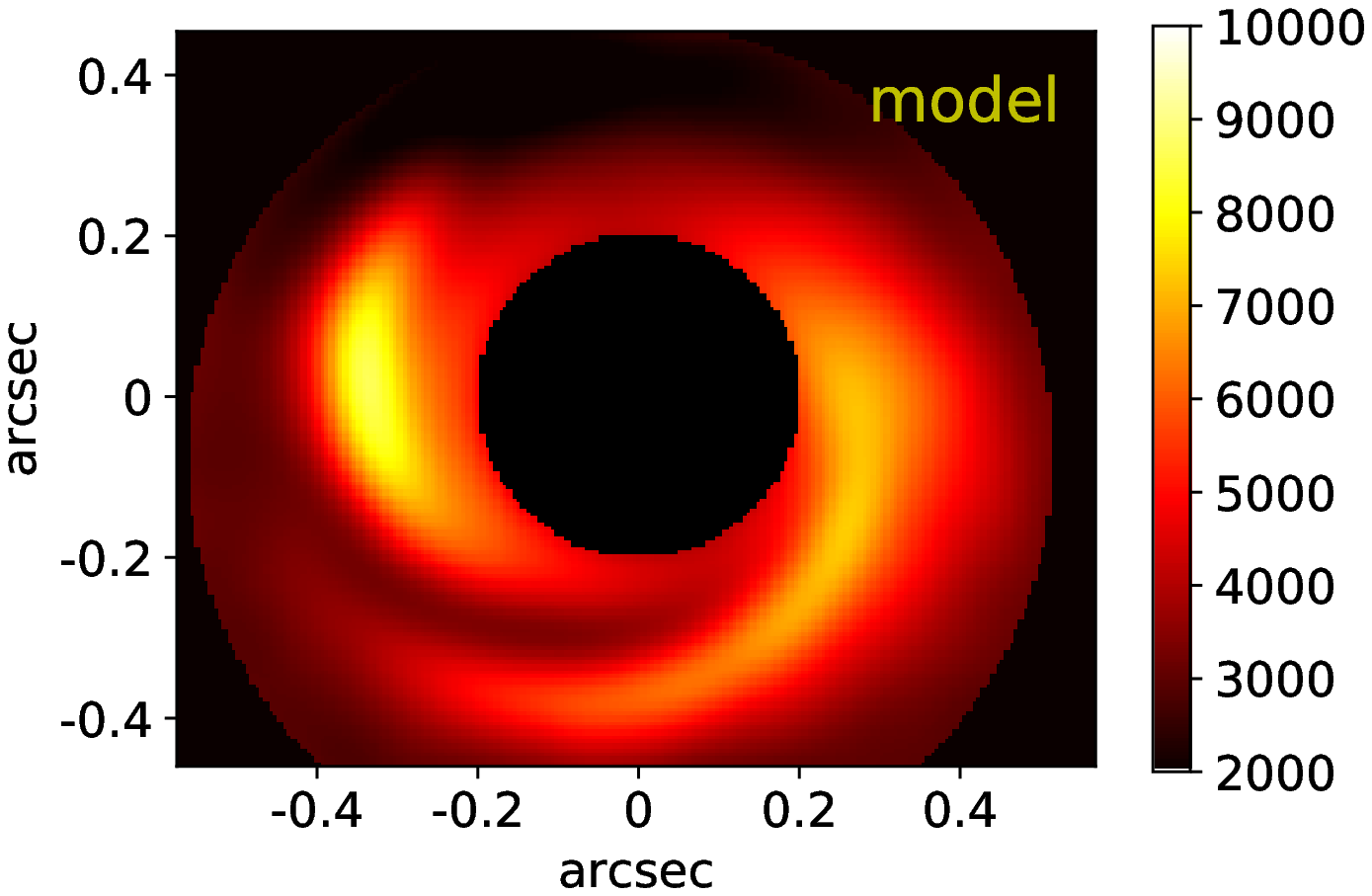}
\end{minipage}
\end{tabular}
\caption{Comparison of $r^2$-scaled surface brightness of outer structures in the $H$-band (left), $K_{\rm s}$-band (middle), and model (right) images. The color bars are arbitrary. We overlay an $r=0\farcs2$ mask on all the images because the model calculation focuses on reproducing only the outer features.}
\label{compare spirals}
\end{figure*}

We carried out three-dimensional hydrodynamics and radiative transfer simulations to produce synthetic images of a pair of planet-induced spiral arms in scattered light that qualitatively match the Subaru observations of LkH$\alpha$ 330, as shown in Figure \ref{compare spirals}. 
The simulation was conducted based on the planetary-mass-companion model in \citet{Dong2016}, and is briefly described here. We use the hydrodynamics code PENGUIN \citep{Fung2015thesis} to calculate the density structure of a disk perturbed by a planet in a circular orbit at 100 au. The resulting disk structure is translated into near-IR polarized light images using the radiative transfer code HO-CHUNK3D \citep{Whitney2013}. The planet's mass was $M_{\rm planet}=0.003M_\star$, which corresponds to $\sim$$6 M_{\rm J}$ in LkH$\alpha$ 330. The synthetic image was produced assuming the object is 310 pc away, and under the actual viewing angle of the system, PA = 90$^\circ$ and inclination $i$ = 30$^\circ$. The image was convolved by a Gaussian PSF to achieve an angular resolution of $0\farcs06$. The outer disk exterior to the planet's orbit was removed in post-processing.
The model image matches the actual data well, which suggests that the two spiral arms in LkH$\alpha$ 330 may be induced by a $5$--$10M_{\rm J}$ planet at $\sim120$ AU ($0\farcs4$). At the current epoch, the planet may be at PA $\sim$20$^\circ$.
Note that these simulations focus on reconstructing a pair of spirals in the LkH$\alpha$ 330's disk and are separate from those reported in \cite{Isella2013} that predicted unseen protoplanets within the gap region.

%We note that the difference in the east direction between the $H$- and $K_{\rm s}$-band images may be caused by a directional shadow, which is independent of the spiral modeling and is discussed in Section \ref{sec: Color Map}.

\subsection{Color Discussion} \label{sec: Color Discussion}
We have observed LkH$\alpha$ 330 in the $H$- and $K_{\rm s}$-bands and can discuss a disk variation in wavelength. In this section we focus on the difference between the azimuthal profile of the ring shown in Figure \ref{azimuthal profile ring} and a color map of the disk.

\subsubsection{Scattering Properties} \label{sec: Scattering Properties}

\begin{figure*}
\centering
\includegraphics[scale=0.48]{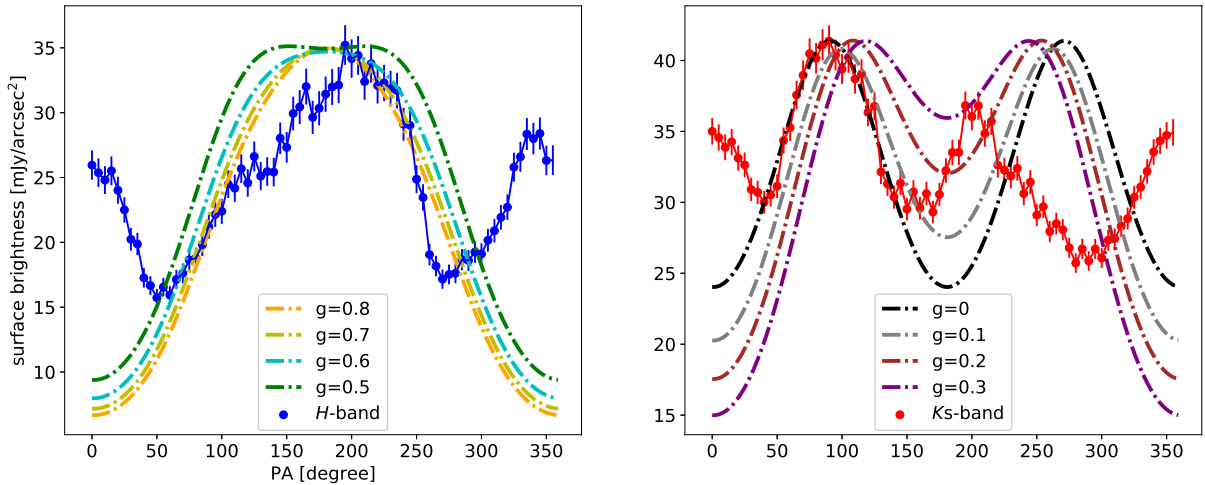}
\caption{Azimuthal surface brightness profiles of the deprojected ring region ($r\sim$$0\farcs17$) in the $H$-band(left) and $K_{\rm s}$-band (right). We include expected phase functions assuming a uniform dust distribution in the disk.}
\label{azimuthal profile ring}
\end{figure*}

\begin{figure}
 \centering
 \includegraphics[scale=0.65]{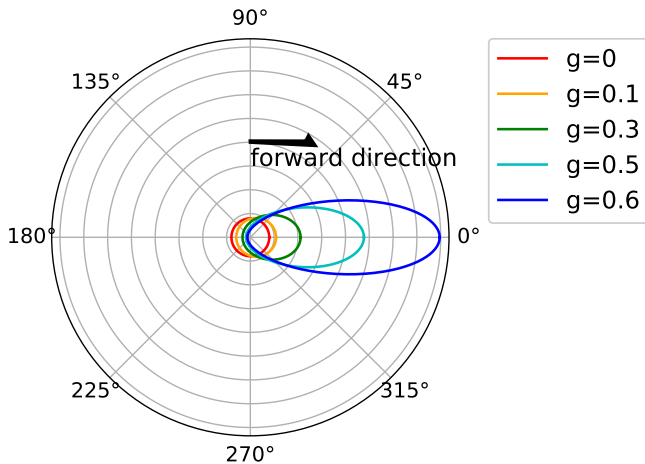}
 \caption{Polar diagram of Henyey-Greenstein phase function. The phase angle corresponds to a deviation from the forward-scattering direction. The radial scale is arbitrary.}
 \label{scattering map}
\end{figure}

We investigated phase functions of a grain-scattering model in Figure \ref{azimuthal profile ring} using equations (8) and (9) in \cite{Graham2007}, where we fixed the polarization parameter of $p_{\rm max}$ to 1.
We put four dotted lines in each graph by changing the scattering parameter of $g$ and a coefficient of the phase function in order to compare the phase functions to the ring profiles.
%Modified by Muto
A formula of the phase function is given by
%as follows:
\begin{eqnarray*}
\propto
\frac{1-g^2}{4\pi(1+g^2-2g\cos{\theta}\sin{i})^{3/2}}\cdot p_{\rm max}\frac{1-\cos^2{\theta}\sin^2{i}}{1+\cos^2{\theta}\sin^2{i}}\ ,
\end{eqnarray*}
where $\theta$ is a scattering angle and $i$ is an inclination.
Figure \ref{scattering map} is a polar diagram of Henyey-Greenstein phase function \citep[$\Phi(a)=\frac{1-g^2}{4\pi(1+g^2-2g\cos(a))^{3/2}}$, where $a$ is a phase angle;][]{Henyey1941} by changing $g$ from 0 to 0.6, which shows the dependency of scattering angle on the phase function. 
Forward scattering angle is given by $\pi/2-i-\beta$ where $\beta$ is an opening angle of the disk \citep[see figure 9 in][]{Jang-Condell2017}. 
We assume that the disk is geometrically thin and $\beta$ is negligible.
Our adopted phase functions partially deviate from the azimuthal profiles of the ring.
These phase functions have similar profiles at the brightest region. 
The $H$-band functions with $g>0.5$ may be more suitable for the profile at PA$\sim$$90^\circ$--$270^\circ$, while $g=0$ fits the profile at PA$\sim$$90^\circ$ in the $K_{\rm s}$-band.
However, these functions do not fully agree with the ring profiles.
Apparently at the north region (PA$\sim$0$^\circ$) both rings are much brighter than the expected phase functions and $K_{\rm s}$-band azimuthal profile three peaks, which cannot be reproduced by a simple phase function we adopt and may suggest the ring has non-axisymmetric distribution of dust and/or composition.

%The overlaid phase functions used $g=0$ and $0.6$ for the $H$- and $K_{\rm s}$-band results, respectively, in order to compare the phase functions to the ring profiles. These phase functions have similar profiles at the brightest region.

%apparently different from phase functions
%need non-axisymmetric pattern - asymmetric density and/or size distribution (ALMA) and/or component (JWST and TMT)

%We have found that very different values of the $g$ parameter to explain the azimuthal profiles of the scattered light on the ring. 
In order to investigate whether such large variations of the scattering properties with a simple dust model, we run a Mie scattering code attached to the HO-CHUNK radiative transfer code \citep{Bohren1983,Whitney2003} assuming the astronomical silicates model \citep[][]{Laor1993}.
When we defined the dust size so as to reproduce the $K_{\rm s}$-band $g=0$ value by changing minimum dust size between 1 nm and 300 nm, the $g$ value in the $H$-band was automatically defined and was always smaller than 0.5.
This result indicates that the ring profiles cannot be described by a simple dust species distributed all over the entire disk.
Our assumptions and calculations could not characterize the disk fully but might set particular constraints on the dust distribution.

Explaining these variations requires non-axisymmetric pattern of size, density, and composition distributions and vertical structures.
As ALMA revealed that MWC 758 has complexity in its ring \citep[][]{Boehler2018}, radio interferometric observations will help to reveal dust size and density distributions. 
%Modified by Muto
Identifying its composition requires follow-up observations in other bands, e.g., with JWST/MIRI \citep[][]{Wells2015} or TMT/MICHI \citep[][]{Sakon2014}. A future integral field spectroscopy (IFS) in mid-IR wavelength can constrain abundance of polycyclic aromatic hydrocarbon and silicate.
The wavelength difference of the azimuthal surface brightness structures 
shown in Figure \ref{azimuthal profile ring}
may not be simply explained by the azimuthal variations of the vertical structures and therefore
%Non-axisymmetric vertical structures may not affect the wavelength-difference and
investigating synthetic dust distributions will be necessary.
%more help.

\subsubsection{Color Map} \label{sec: Color Map}
\begin{figure}
 \centering
 \includegraphics[scale=0.8]{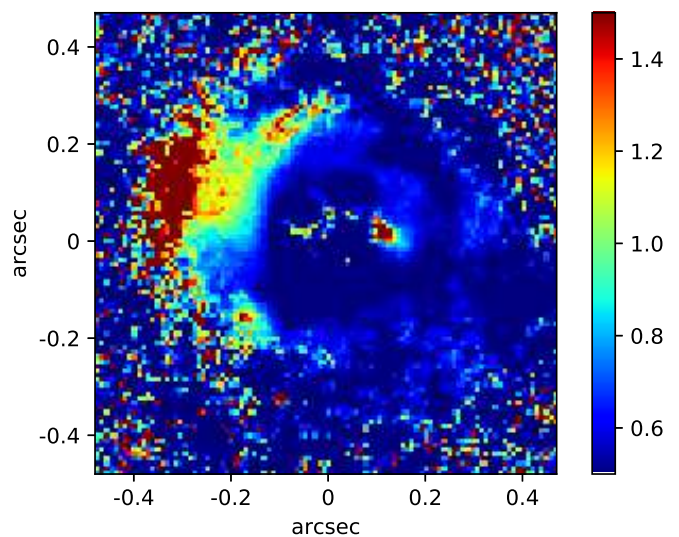}
 \caption{Color map of the LkH$\alpha$ 330 disk. The scale bar represents the PI ratios $K_{\rm s}/H$ such that red represents a larger value.}
 \label{color map}
\end{figure}

Figure \ref{color map} shows a color map of the disk generated by dividing the $K_{\rm s}$-band PI image by the $H$-band PI image. In this process the $K_{\rm s}$-band image is convolved in order to fit the $K_{\rm s}$-band's PSF to the $H$-band's PSF. 
AO188 worked effectively enough to suppress both PSF's wings and the $H$-band PSF's core is broader as mentioned in Section \ref{sec: Observations and Data Reduction}.
We then re-registered two PDI-reduced images by defining the center as the elliptical-fit results of the ring.
Finally we normalized the $K_{\rm s}$/$H$ PI image by the $K_{\rm s}/H$ luminosity of the central star \citep[=2.27 from UCAC4 catalog;][]{Zacharias2012}.

We find that the disk is basically ``blue".
This implies that the typical dust size is small enough for Rayleigh scattering. 
However, there are seen some "redder" regions at ($\rho$,PA)=($\sim$$0\farcs3$, 0$^\circ$--30$^\circ$) and ($0\farcs15$--$0\farcs4$, 45$^\circ$--90$^\circ$).
The northern part ($\sim$$0\farcs3$, $\sim$0$^\circ$--30$^\circ$) is influenced by the north-east spiral-like feature that appears only in the $H$-band image (see Figures \ref{spiral fit}, \ref{r-theta H}, and \ref{r-theta K}).
The inner eastern part ($0\farcs15$--$0\farcs2$, 45$^\circ$--90$^\circ$) corresponds to the wavelength-difference of the azimuthal profiles, which is discussed in Section \ref{sec: Scattering Properties}.

The outer eastern part ($0\farcs2$--$0\farcs4$, 45$^\circ$--90$^\circ$) comes from the lack of an outer scattering area in the $H$-band disk. 
The $H$-band PI image previously reported in \citep{Akiyama2016}, which is presented as a $r^2$-scaled PI image in Figure \ref{old H}, looks like the $K_{\rm s}$-band image. A comparison of the FWHMs of the previous and our $H$-band image shows that our data sets had better AO efficiency. Therefore our data reduction shows a ``red" region at the outer north-east region. 
This phenomenon demonstrates that a possibility of ``a directional shadow" is plausible; namely, when we observed this system, an inner object occasionally partially veiled starlight and cast a shadow onto the north-east region of the disk. 
We consider a clump-like object moving in the very inner region. If the orbit of the object is highly misaligned with the outer disk, we can explain the change of outer morphology over the 3 months. The orbital separation is expected to be very near ($\leq$1 au) the central star. Regarding the possibility of there being an inner clump, we refer to CVSO 30 and ``dipper" stars. CVSO 30 was reported to harbor a close-in protoplanet candidate based on transit observations \citep[CVSO 30 b;][]{vanEyken2012}. However, follow-up observations suggested that the companion candidate is a clump rather than a planet \citep[][]{Onitsuka2017}. Dipper YSOs revealed by the K2 survey suggest the existence of occulting structures located at quite small separations \citep[][]{Ansdell2016}. Particularly, \cite{Ansdell2016} reported RX J1604.3-2130A to be a dipper star. This YSO has a face-on disk with a large gap \citep[e.g.,][]{Mayama2012}. Therefore an inner clump-like object with an orbit highly misaligned to the outer disk is a possible scenario.

%Modified by Muto
Another possible mechanism of casting shadow in the outer disk is the existence of an inner disk.
%We also take into account of the possibility of an inner disk.
Previous disk observations have revealed shadows on the protoplanetary disks induced by inner objects \citep[][]{Garufi2014,Pinilla2015,Canovas2017,Stolker2016,Stolker2017,Bensity2017,Debes2017}.
%Thus if 
If LkH$\alpha$ 330 has an inner disk that can cast shadows on the outer disk, non-axisymmetric structures can be observed \citep[][]{Facchini2018}.
From the models presented in \cite{Long2017}, we consider that the difference of the inclination between the inner and the outer disks is as small as $<10^{\circ}$ in order to produce the azimuthal features observed in $H$-band.
%ここのロジックが、ちょっと追いにくい。Hバンドを説明するためには、外側と内側の円盤の inclination のずれが 10度以下程度であろう、ということ？次の文章の「such a feature」が何を表しているか？
%A small difference of 
%inclination ($<10^\circ$) between the inner and outer disks can describe this feature \citep[][]{Long2017}.
Given that our $K_{\rm s}$-band image, which was taken just 3 months after the $H$-band image, however, 
%has different outer shape, %after just 3 months,  %does not show such a feature.
this rapid change of the shadow feature can hardly be explained by the change of the orientation of the inner disk because we do not expect rapid (on the timescale comparable to Kepler timescale) precession of the inner disk.
%Otherwise the feature should be steady and 
The inner disk should have a particular extinction property to let $K_{\rm s}$-band wavelength transmit.
As multi-band observations of HD 100453 \citep[][]{Bensity2017} did not show clear difference of the shadow structures, we consider that the shadowing by the inner disk is unlikely.
%the extinction property of the inner disk may be an invalid scenario.

In order to investigate this scenario follow-up observations are required. If the PI signal is detected again as reported in \cite{Akiyama2016}, a clump-like object scenario is plausible. However, if follow-up observations fail to detect a PI signal, 
%another scenario 
%Added by Muto
other scenarios should be considered, such as asymmetric dust distribution or difference of dust properties between the outer and inner disks.

%\subsubsection{Disk Composition Scenario}
%Assuming that the possibility of the directional shadow by an inner clump is denied, explaining the color map requires dust characteristics at outer east region.
%If the disk's surface at this region has dust whose size is $\sim$0.15--0.2$\mu$m, $H$-band wavelength experiences Mie scattering while $K_{\rm s}$ scatters as Rayleigh scattering rather than Mie scattering.
%Given that the rest of the regions are blue and hold smaller dust at their surfaces, the redder region corresponds to a location where grain growth is proceeding the most.

%Our data sets show that LkH$\alpha$ 330 is an intriguing system with a gap and spiral-like structures around a T Tauri Star. This object can be a good laboratory for investigating planet formation, disk evolution, and inner structures and is one of the best candidates for follow-up observations at mid-IR wavelengths as well as near-IR, mm and sub-mm wavelengths.

\acknowledgments
The authors thank David Lafreni$\grave{e}$re for generously providing the source code for the LOCI algorithm.
The authors would like to thank the anonymous referees for their constructive comments and suggestions to improve the quality of the paper.
This research is based on data collected at the Subaru Telescope, which is operated by the National Astronomical Observatories of Japan.
Based in part on data collected at Subaru telescope and obtained from the SMOKA, which is operated by the Astronomy Data Center, National Astronomical Observatory of Japan.
Data analysis were carried out on common use data analysis computer system at the Astronomy Data Center, ADC, of the National Astronomical Observatory of Japan. 
This research has made use of NASA's Astrophysics Data System Bibliographic Services.
This research has made use of the SIMBAD database, operated at CDS, Strasbourg, France. 
This research has made use of the VizieR catalogue access tool, CDS, Strasbourg, France. The original description of the VizieR service was published in A{\&}AS 143, 23.

This work was supported by JSPS KAKENHI Grant Numbers JP17J00934, JP15H02063, JP258826.

The authors wish to acknowledge the very significant cultural role and reverence that the summit of Mauna Kea has always had within the indigenous Hawaiian community. We are most fortunate to have the opportunity to conduct observations from this mountain.

\bibliographystyle{apj}                                                              %use bibtex
\bibliography{library}                                                                %use bibtex

%\clearpage
\appendix
\section{Error Map of Polarization Angle}

\begin{figure*}[h]
\begin{tabular}{cc}
\begin{minipage}{0.5\hsize}
 \centering
 \includegraphics[scale=0.65]{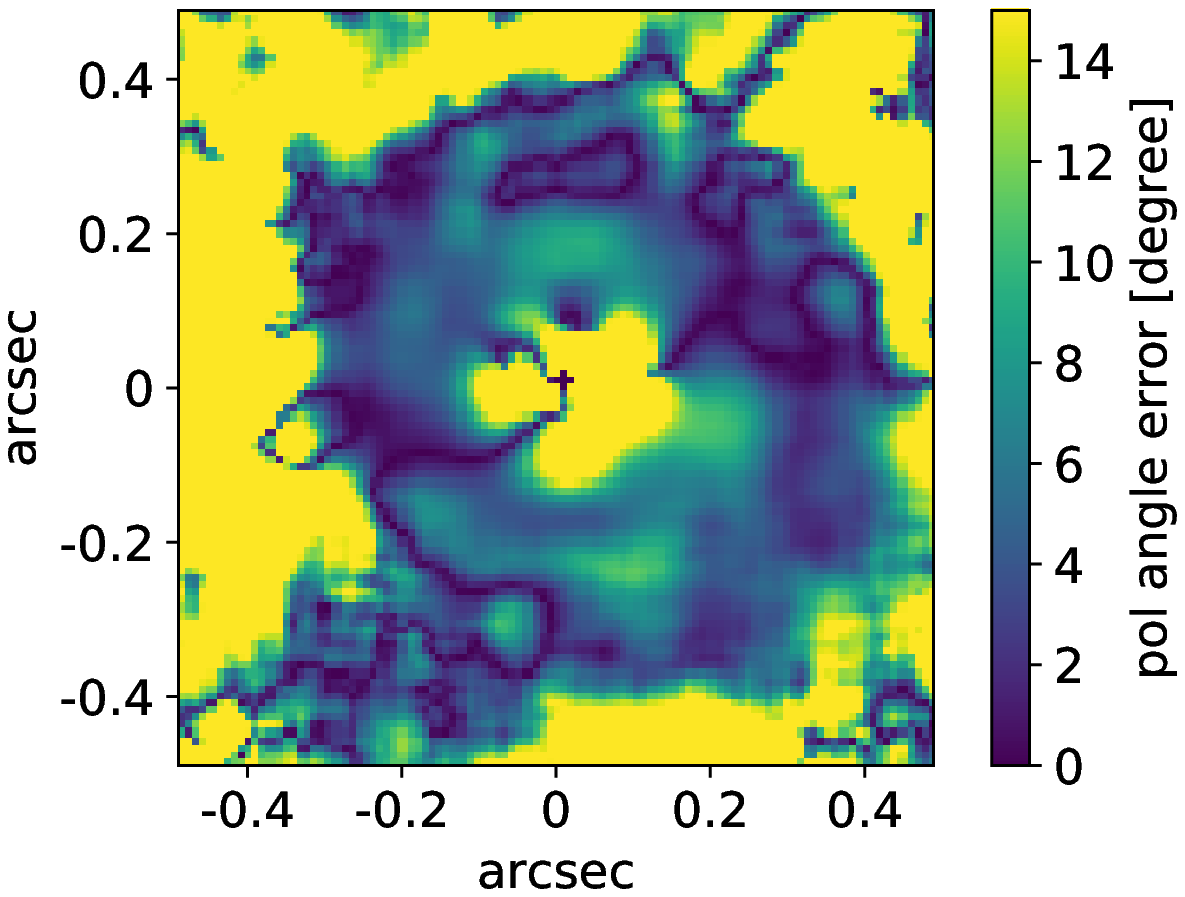}
\end{minipage}
\begin{minipage}{0.5\hsize}
 \centering
 \includegraphics[scale=0.65]{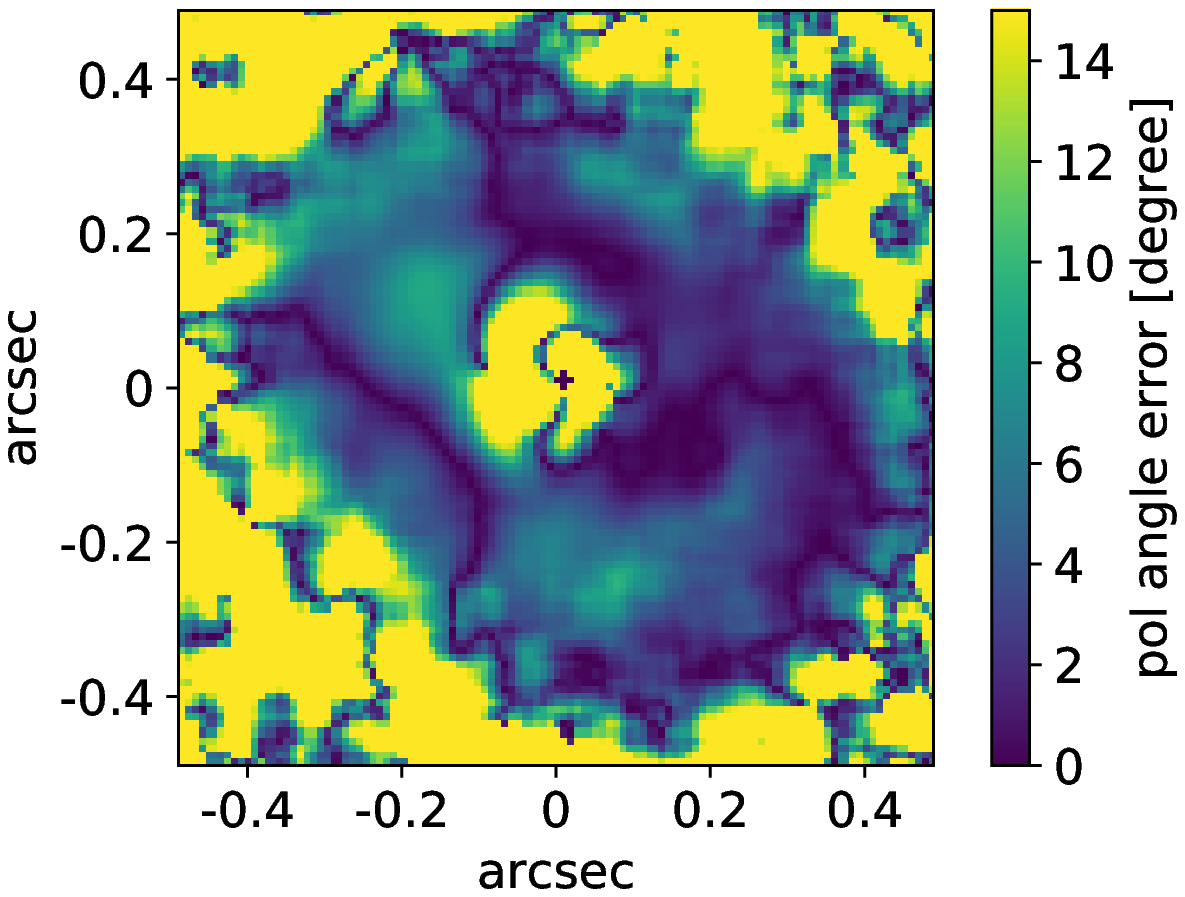}
\end{minipage}
\end{tabular}
\caption{Angle error maps of the $H$- (left) and $K_{\rm s}$-band (right) images. These maps are used for calculating radial noise of the PI images, which is described in Section \ref{sec: Results and Data Analysis}.}
\label{angle error map}
\end{figure*}

Figure \ref{angle error map}, which is made from Figure \ref{pol vector}, shows angle error maps in the $H$- and $K_{\rm s}$-bands.
These maps are used for calculating noise profiles of the PI images and details are explained in Section \ref{sec: Results and Data Analysis}.
Yellow corresponds to larger difference from a centro-symmetric pattern of the polarization vectors and causes larger noise in the PI image.

\section{Comparison of Our results with the Previous HiCIAO Image}
\begin{figure*}[h]
\begin{tabular}{ccc}
\begin{minipage}{0.33\hsize}
 \centering
 \includegraphics[scale=0.5]{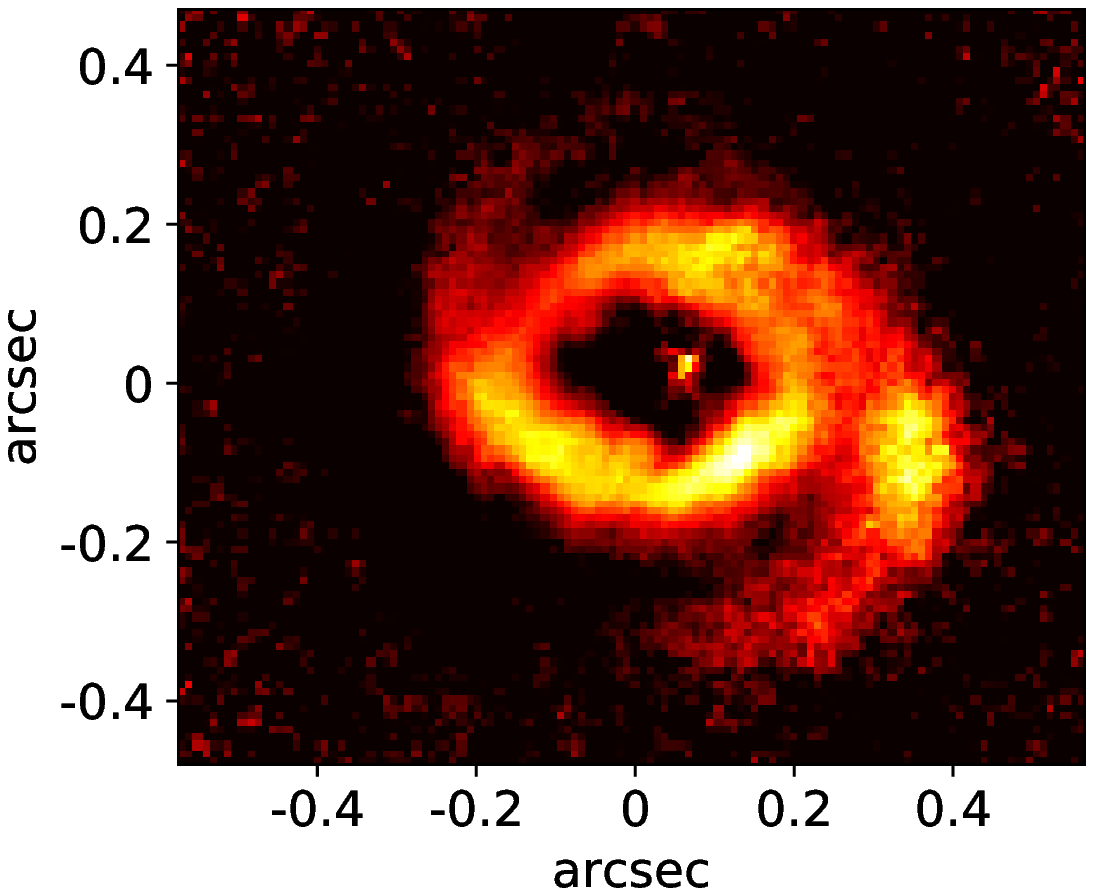}
\end{minipage}
\begin{minipage}{0.33\hsize}
 \centering
 \includegraphics[scale=0.5]{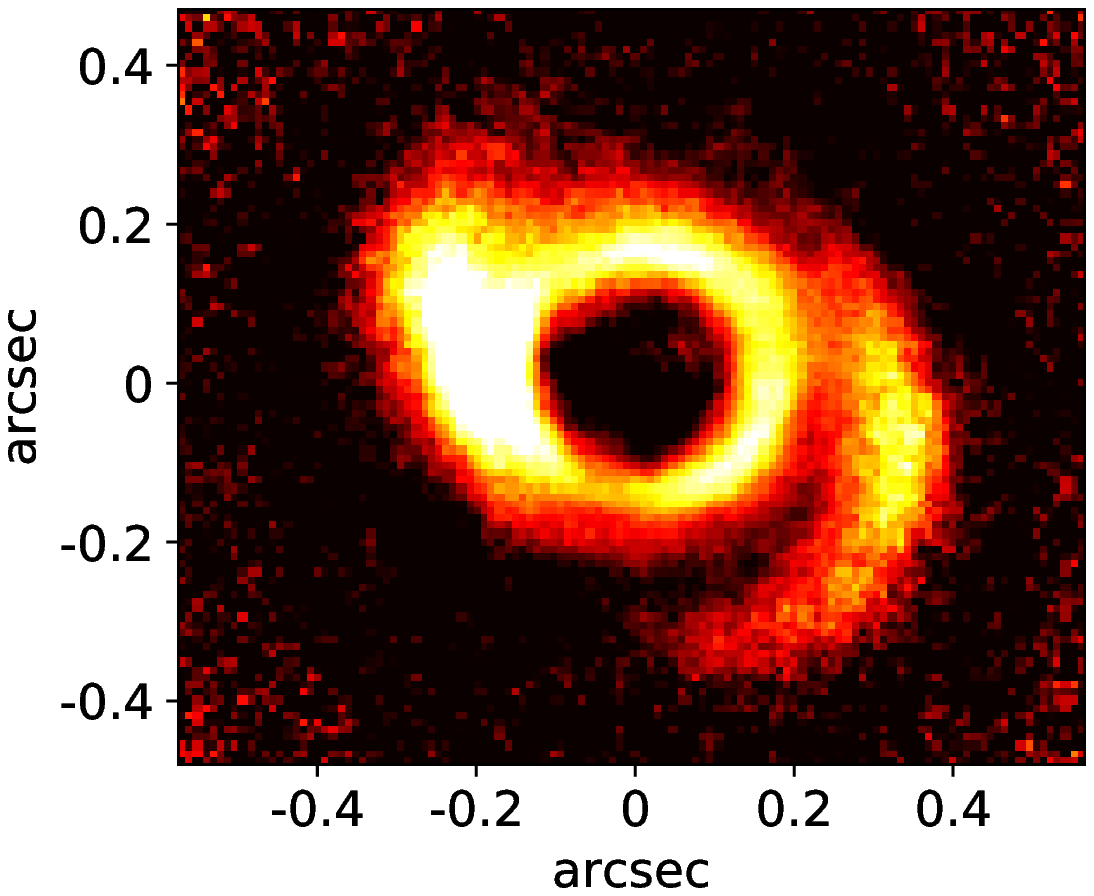}
\end{minipage}
\begin{minipage}{0.33\hsize}
 \centering
 \includegraphics[scale=0.5]{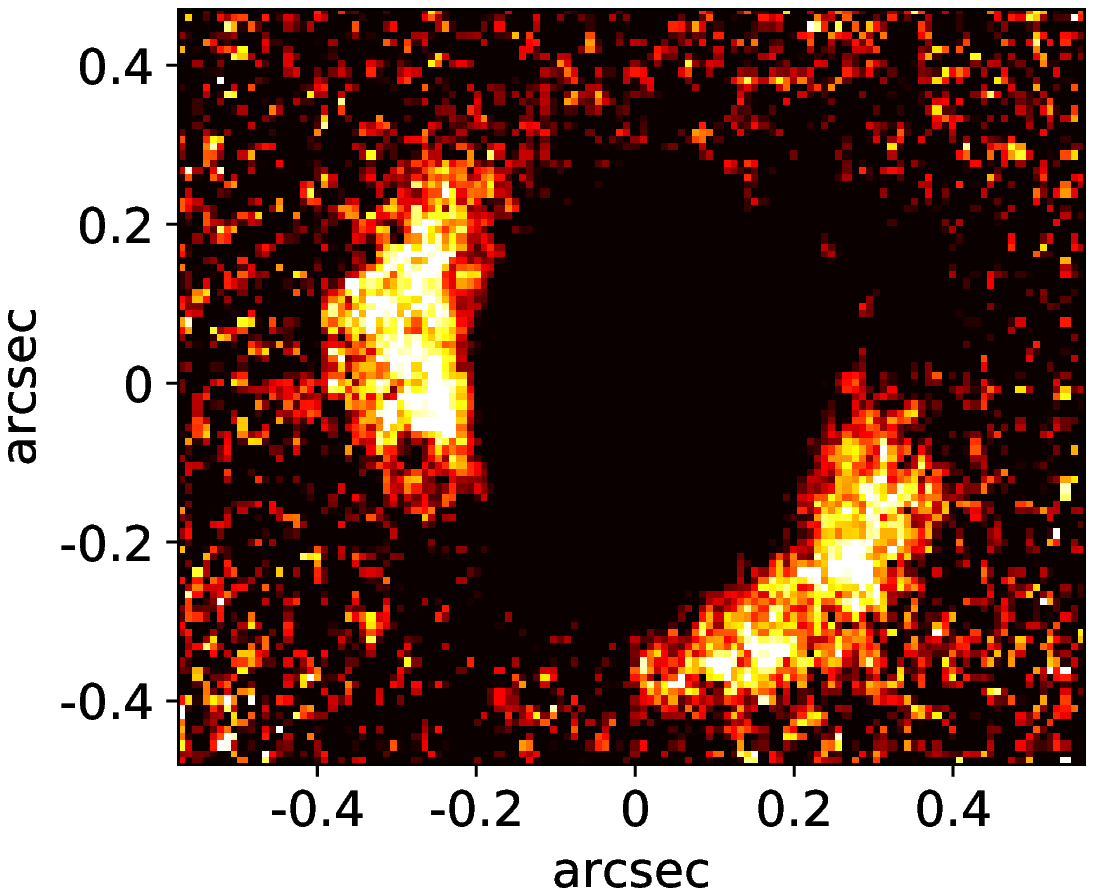}
\end{minipage}
\end{tabular}
\caption{$r^2$-scaled PI images in the $H$-band taken in 2014 (left), the $K_{\rm s}$-band taken in 2015 (center), and the $H$-band taken in 2011 (right). The right image is masked with a $0\farcs2$-radius mask.}
\label{old H}
\end{figure*}

Figure \ref{old H} compares our HiCIAO observations with the previous HiCIAO observation that was originally presented in \cite{Akiyama2016}. Our data sets achieved better AO efficiency and are explained in Section \ref{sec: Observations and Data Reduction}.

\end{document}